%% file: 47Tuc_paper1_v11.tex
\documentclass{aa}
\usepackage{graphicx}
\usepackage{txfonts}
\usepackage{epsf}  
\usepackage{amssymb}  
\usepackage{rotating}
\usepackage{lscape} 
\usepackage{multirow}
\usepackage{url}
\usepackage{longtable}
\usepackage{natbib}
\bibpunct{(}{)}{;}{a}{}{,} 
\begin{document}

%
%
\newcommand{\ie}{i.e.}
\newcommand{\eg}{e.g.}
\newcommand{\cf}{cf.}	
\newcommand{\kms}{km\,s$^{-1}$}
\newcommand{\teff}{$T_{\rm eff}$}
\newcommand{\logg}{$\log g$}
\newcommand{\feh}{[Fe/H]}
\newcommand{\msun}{${\rm M}_{\odot}$}
\newcommand{\percent}{\,{\%}}
\newcommand{\vmic}{$\xi_t$}
\newcommand{\vsini}{$v \sin i$}
\newcommand{\feone}{Fe\,{\sc I}}
\newcommand{\fetwo}{Fe\,{\sc II}}
\newcommand{\loggf}{log$(gf)$}

\title{The chemical composition of red giants in 47 Tucanae I: Fundamental parameters and chemical abundance patterns
\thanks{Based on observations made with the ESO Very Large Telescope at Paranal Observatory, Chile (Programmes 084.B-0810 and 086.B-0237).Tables 2, 5, 8 and 9 are available in electronic form at the CDS via anonymous ftp to cdsarc.u-strasbg.fr (130.79.128.5) or via http://cdsweb.u-strasbg.fr/cgi-bin/qcat?J/A+A/.}}
\titlerunning{The chemical composition of 47Tuc I}
\authorrunning{A. O. Thygesen et al.}
\author{
A. O.~Thygesen\inst{1} 
\and
L.~Sbordone\inst{1,2,3}
\and
S.~Andrievsky\inst{4,5}
\and
S.~Korotin\inst{4}
\and
D.~Yong\inst{6}
\and
S.~Zaggia\inst{7}
\and
H.-G.~Ludwig\inst{1}
\and
R.~Collet\inst{6}
\and
M.~Asplund\inst{6}
\and
P.~Ventura\inst{8}
\and
F.~D'Antona\inst{8}
\and
J.~Mel{\'e}ndez\inst{9}
\and
A.~D'Ercole\inst{10}
} 
\offprints{A. O.~Thygesen}
\mail{a.thygesen@lsw.uni-heidelberg.de}
\institute{Zentrum f\"{u}r Astronomie der Universit\"{a}t Heidelberg, Landessternwarte, K\"{o}nigstuhl 12, 69117 Heidelberg, Germany.
\and
Millennium Institute of Astrophysics, Av. Vicu{\~n}a Mackenna 4860, 782-0436 Macul, Santiago, Chile.
\and
Pontificia Universidad Cat{\'o}lica de Chile, Av. Vicu{\~n}a Mackenna 4860, 782-0436 Macul, Santiago, Chile.
\and
Department of Astronomy and Astronomical Observatory, Odessa National University, Isaac Newton Institute of Chile, Odessa Branch, 
Shevchenko Park, 65014, Odessa, Ukraine.
\and
GEPI, Observatoire de Paris, CNRS, Université Paris Diderot, Place Jules Janssen, 92190, Meudon, France.
\and
Research School of Astronomy and Astrophysics, Australian National University, Canberra, ACT 0200, Australia.
\and
INAF-Osservatorio Astronomico di Padova, Vicolo dell'Osservatorio 5, 35122 Padova, Italy.
\and
INAF-Osservatorio Astronomico di Roma, Via Frascati 33, I-00040 Monte Porzio Catone, Italy.
\and
Departamento de Astronomia do IAG/USP, Universidade de São Paulo, rua do Mãtao 1226, 05508-900, São Paulo, SP, Brasil.
\and
INAF-Osservatorio Astronomico di Bologna, Via Ranzani 1, I-40127 Bologna, Italy.
}
\date{Received 04 July 2014 ; Accepted 12 September 2014 }
\abstract
{The study of chemical abundance patterns in globular clusters is of key importance to constrain the different candidates for intra-cluster pollution of light elements.} 
{We aim at deriving accurate abundances for a large range of elements in the globular cluster 47 Tucanae (NGC 104) to add new constraints to the pollution scenarios for this particular cluster, expanding the range of previously derived element abundances.}
{Using tailored 1D LTE atmospheric models together with a combination of equivalent width measurements, LTE, and NLTE synthesis we derive stellar parameters and element abundances from high-resolution, high signal-to-noise spectra of 13 red giant stars near the tip of the RGB.}
{We derive abundances of a total 27 elements (O, Na, Mg, Al, Si, Ca, Sc, Ti, V, Cr, Mn, Fe, Co, Ni, Cu, Zn, Y, Zr, Mo, Ru, Ba, La, Ce, Pr, Nd, Eu, Dy). Departures from LTE were taken into account for Na, Al and Ba. We find a mean [Fe/H] = $-0.78\pm0.07$ and $[\alpha/{\rm Fe}]=0.34\pm0.03$ in good agreement with previous studies. The remaining elements show good agreement with the literature, but the inclusion of NLTE for Al has a significant impact on the behaviour of this key element.}
{We confirm the presence of an Na-O anti-correlation in 47 Tucanae found by several other works. Our NLTE analysis of Al shifts the [Al/Fe] to lower values, indicating that this may be overestimated in earlier works. No evidence for an intrinsic variation is found in any of the remaining elements.}
\maketitle

\section{Introduction}

Over the last few decades it has been realised that most, if not all, globular clusters (GCs) host at least two populations of stars, thus removing their previous status as single stellar populations. This has spawned a renewed interest in GC studies, in order to understand the underlying cause of the multiple populations observed. These populations reveal themselves by splittings of the main sequence (MS), turn-off (TO), sub-giant branch (SGB) or red giant branch (RGB) in high-quality photometry (\eg\ \citealt{milone,piotto,piotto2,monelli}) as well as in a variation of light elements (He, CNO, Na, Mg, Al) across the stellar populations (\eg\ \citealt{yong2,carretta2,ivans,kacharov}). 

In a few exceptional cases, also a variation in Fe is observed, e.g. in $\omega$~Centauri (\citealt{norris,smith,johnson}) or M22 \citep{marino,dacosta}. But the vast majority of clusters do not show any star-to-star variation in the iron-peak elements. A particularly striking feature in the light element abundance variations is an observed anti-correlation between Na and O, present in essentially all GCs, and a weaker anti-correlation between Mg and Al in some cases. These stars, which show variations in the light elements, different from field stars at the same metallicity are commonly referred to as polluted or enriched stars.The presence of these multiple populations is in fact so common that it has been suggested as a defining feature of GCs. For a recent review we refer the interested reader to \citet{gratton} and references therein.

It has long been known that differences in the strengths of CN and CH molecular features can affect the observed colours of the stars \citep{bond}. In the context of light element variations in globular clusters, \citet{grundahl2} showed that the N variations could be traced using Str{\"o}mgren photometry. Also \citet{yong3} and \citet{marino2} found that Na-rich/poor stars could be identified from particular photometric colour combinations. The full connection between abundance variations and photometry was, however, only fully explained through the theoretical modelling of the spectra by \citet{sbordone2} and was further explored through a study of a number of different photometric indices by \citet{carrettaphot}. Their studies showed that the polluted generations of stars were typically found to be bluer than their pristine counterparts in the cluster colour-magnitude diagrams (CMDs), when observed in (B-V), but whether the polluted population appears more red or more blue than their pristine counterpart, depends on the exact photometric filter combination used.

In particular the mechanism responsible for the light element variation is a matter of active research. The variation is commonly agreed to have been caused by self-pollution within the cluster, rather than being linked to the evolutionary state of the stars, since the variations are observed in stars of all stages of evolution, including MS stars. Stars currently on the MS of GCs are of too low mass to reach the temperatures required for nuclear burnings advanced enough to produce the observed abundance variations (\ie\,25MK for the Ne-Na burning chain, and 70\,MK for Mg-Al, \citealt{denisenkov,langer,prantzos}). The ejecta from supernovae is usually not considered to have contributed to the enrichment due to the mono-metallicity observed in most clusters. However, the study by \citet{marcolini} indeed suggests that a combination of type II and type Ia supernovae can also explain the observed abundance variations in some cases.

One of the most widely accepted explanations for the intra-cluster pollution is an early generation of stars of  
intermediate mass during the AGB phase \citep{ventura2009}. These stars are able to undergo the required nuclear burnings as well as release the enriched gas in thermal pulses, where the shedded material has a velocity low enough that it can be retained within the gravitational potential well of the cluster \citep{dercole,ventura2,dercole2}. This scenario, however, requires the GCs to have been $\sim$10 times more massive in the past in order to reproduce the approximate 1:1 ratio between members of the pristine and polluted populations of stars. This assumption was recently challenged by \citet{larsen}, who found that GCs in the Fornax dwarf galaxy could have been at most five times more massive in the past. Further, the predicted AGB yields are in some cases at odds with what is derived from observations \citep{fenner,bekki} and substantial disagreement between the yields from different models are still seen in the literature \citep{denissenkovAGB,karakas2,doherty,ventura}. Also, for the AGB scenario to be successful, only a narrow mass range should contribute to the intra-cluster gas, requiring a non-standard initial mass function (IMF).

Another popular polluter candidate is Very Massive Fast Rotators \citep{decressin}, which also release light elements into the cluster environment through a slow wind at an early point in the cluster evolution. However, this scenario also requires an anomalous IMF in order to yield sufficient amounts of enriched gas. In addition, the activation of the Mg-Al cycle in the Very Massive Fast Rotator models would require the currently used cross section for proton capture on $^{24}$Mg to be underestimated by 2-3 orders of magnitude \citep{decressin}. A variation of this was proposed by \citet{demink}, who focused on the role of massive, interacting binaries as source of the abundance variations. Interacting binaries provide a very efficient mechanism for mass-loss, not requiring an anomalous IMF in order to produce enough enriched material. However, the latter scenario relies on the assumptions that the IMF of the polluted generation truncates at $\sim0.8$\,\msun\ and that all massive stars in the cluster are in interacting binary systems.

Recently, two additional scenarios were proposed. \citet{bastian} propose that low-mass proto-stars accrete enriched material before settling on the main sequence, again providing a mechanism for creating an enriched population of stars, a scenario where pollution could be achieved without a non-standard IMF nor requiring the cluster to have been substantially more massive in the past. It does, however, rely on a number of assumptions about, \eg,  the effectivenes of accretion on proto-stars and the survival time of the proto-stellar discs in a dense environment like a GC, which needs further detailed work to be fully applied to GCs. 

The study of \citet{denissenkov} instead focuses on the role of Super Massive Stars ($\sim10^4$\,\msun) as polluters, although only in an explorative way. More work is needed for both these candidates to properly test them against observations. In summary, all proposed candidates are able to explain parts of the observed behaviour in GCs, but the complete observational picture still lacks a coherent explanation.  

The GC 47 Tucanae (NGC104) is one of the brightest GCs in the sky, with an apparent V magnitude of 4.09 \citep{dalessandro}, rivaled only in brightness by $\omega$~Centauri. It is also one of the most massive GCs in the Milky Way, with a total mass of $7\times10^5$\msun\ \citep{marks}, and furthermore it is in the metal-rich end of the GC population. Its close distance makes it one of the most well-studied clusters in our Galaxy. Photometric studies during the last decade, have revealed that 47 Tuc contains multiple stellar populations. Two SGB populations were discovered by \citet{anderson}, who also noted a clear broadening of the MS, but were unable to identify distinct populations. More recently, \citet{milone} were also able to identify two populations, identifying them on the MS as well as on the SGB, RGB and horizontal branch. They found that $\sim$30\% of the stars currently observed in the cluster belong to the pristine population, so that most stars will have been polluted to some extent. They also identified a third population of stars, making up only $\sim$8\% of the cluster stars and visible only on the SGB.

Additional evidence for multiple populations was recently put forward by \citet{richer}, who identified different proper motion anisotropies between the blue and the red MS stars, being particularly evident for the blue sequence. The stars belonging to the bluer sequence were also found to be more centrally concentrated, in accordance with the findings of \citet{milone}.

A large number of spectroscopic abundance studies of 47 Tuc are available already in the literature, dating back to the work of \citet{dickens}, who noted a variation in the nitrogen abundance, which was later confirmed by \citet{norris2,hesser} and \citet{cottrell} to name a few. Other studies also found evidence for variations in the light element abundances, \eg\ \citet{briley}, who identified variations in Na, and found CN and CH to be anti-correlated in a number of stars at the main sequence turn-off, putting stringent limits on mixing scenarios as the cause for the variations. More recent studies have reported variations also in Na and O \citep{briley2,koch,alves-brito,dobrovolskas,cordero}. The latter three studies also reported variations in Mg and/or Al. Recently \citet{ventura} proposed a pollution scenario for 47 Tuc under the AGB scheme, able to explain the previously observed variation in Na, O and Al, assuming a degree of dilution with pristine gas within the cluster. However, to reproduce the distribution of pristine vs. polluted stars, the study does require the cluster to have been $\sim7.5$ times more massive in the past, at odds with the constraints from the \citet{larsen} study, but the clusters will likely have had different formation paths, considering the differences in mass (total mass in the 4 Fornax GCs $\sim$$1\times10^6$\,\msun). Also, the IMF of the polluted population of stars needs to be truncated at 5\,\msun. In addition, indications of a variation of S was found by \citet{sbordone}, using high-resolution spectra. Numerous other studies have also reported abundances for a larger range of elements \eg, \citet{brown,carretta2004,alves-brito,mcwilliam} and \citet{dorazi}, to name a few. 

All in all there is ample evidence for multiple stellar populations in 47 Tucanae. Here, we expand the range of derived abundances, by analyzing 13 bright giants, covering 27 different elements derived from the highest quality spectra (in terms of spectral resolution and signal-to-noise). From this dataset we conduct a comprehensive study of abundances in 47 Tuc, adding measurements of Ru, Ce, Pr and Dy to the abundance pattern. Elements which, to the best of the authors knowledge, have not previously been measured in this globular cluster. The measurements will provide additional insight into the chemical evolution history of the cluster, and help constrain the formation history of 47 Tucanae. This work represents the first part of a project aimed at deriving accurate magnesium isotopic ratios in the same stars (Thygesen et al., in preparation; hereafter Paper II).. 

\section{Observations and reduction}

\begin{figure}%
\centering
\includegraphics[width=\columnwidth, trim= 0cm 5cm 0cm 2.5cm, clip=true]{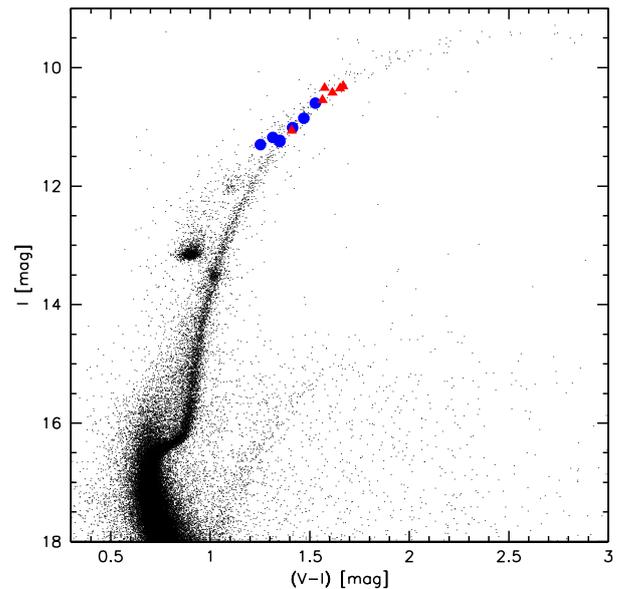}%
\caption{Color-magnitude diagram of 47 Tucanae with observed targets indicated. Blue circles and red triangles are used for the pristine and polluted population of stars respectively. See Sect.~\ref{polluters} for population selection criteria.}%
\label{47tuc_cmd}%
\end{figure}

For this project we observed a total of 13 red giants. The target selection was performed from UBVI photometry obtained from the reduction of archival images taken at the WFI imager at the MPG/ESO 2.2m telescope in La Silla in 2002 under programme 69.D-0582(A). The sample of images comprise short and long time exposures carefully chosen in order not to saturate the brighter targets. The data reduction has been performed using the ESO/MVM pipeline \citep{vandame}, properly correcting for the sky illumination, and fully astrometrizing the whole dataset. Point-spread function photometry has been extracted using DAOPHOT \citep{stetson} and combined in a single photometric dataset. Finally, the sample has been calibrated using the photometric fields compiled by P. Stetson\footnote{\url{http://www3.cadc-ccda.hia-iha.nrc-cnrc.gc.ca/community/STETSON/standards/}}, located inside the WFI field of view of 47 Tuc. Typical rms of the photometry is under 3\% for each UBVI band. The target stars were chosen to be bright (V$\lesssim12.6$), in order to reach the desired signal-to-noise (S/N $\approx150@5140$\AA) in a reasonable amount of time. 47 Tucanae has a very well populated RGB and by restricting the selection to a narrow range of V-magnitudes, we expected to include targets across both populations present in the cluster. However, as is evident from Fig.~\ref{47tuc_cmd}, at the magnitudes of our stars, there is no clear separation between the two populations, making it difficult to guarantee an even sampling across the populations from photometry alone (see also \citealt{milone}). Already published radial velocity measurements were also used in the selection process to confirm cluster membership. We separate the two populations in the CMD, using blue circles and red triangles for the pristine and polluted population of stars respectively, as identified from the spectroscopic analysis (see Sect.~\ref{polluters}).

All spectroscopic observations for this project were aquired under ESO programmes 084.B-0810 and 086.B-0237, using the UVES spectrograph \citep{dekker} mounted on UT2 of the ESO VLT on Cerro Paranal, Chile. Since the main goal of the observations was measurements of the isotopic mixture of Mg from the MgH molecular bands, we required both high signal-to-noise (S/N) as well as high resolution, as the signatures of the heavy Mg isotopes are very subtle. In order to achieve the required S/N without compromising the resolution, we used the image slicer, IS\# 3, reaching a resolving power of $R=110,000$. We used the 580\,nm setting, which covers the wavelength range from 4780\,\AA\ to 6810\,\AA, with a small gap from 5760-5830\,\AA. This setting was chosen in order to capture the MgH bands around 5130\,\AA, which are usually adopted for analyses of Mg-isotopic ratios. An overview of our targets is given in Table~\ref{targetproperties}. The targets were selected to have \logg\,$\leq1.5$ to be sufficiently bright, and \teff\,$<5000$\,K, as the MgH bands largely vanish above this temperature. 

The spectroscopic data was reduced with the ESO GASGANO pipeline V.2.4.3 \citep{silva}\footnote{Available from \url{http://www.eso.org/sci/software/gasgano.html}}. The pipeline performs the standard reduction steps of bias subtraction, flat-field correction, order extraction and wavelength calibration of the observed spectra. Each reduction produces two spectra, one for each chip on the detector. Contrary to the normal procedure of the reduction of echelle spectra, we did not use the optimal extraction method, but a simpler, average extraction. Optimal extraction cannot be applied to image-slicer spectra, as the cross-order profile of the sliced spectrum is unknown and hence very difficult to model. Also, as the slices fill the individual orders completely, it is not possible to do sky subtraction during the reduction, since there is not enough space to determine the sky background accurately. 

After the reduction, the spectra were shifted to laboratory wavelength. First, by correcting for heliocentric motion and subsequently correcting for the radial velocities (RV) of the targets. The velocity shifts have been determined using cross-correlation functions (CCF). The spectra for each detector chip were cross-correlated against a synthetic template spectrum with parameters close to what would be expected for our stars (\teff\ $=4200$K, \logg\ $=1.5$dex, \vmic\ $=2$\kms, \feh\ $=0.00$ dex) and the resulting CCF fitted with a Gaussian to determine the radial velocity of the stars. The quoted uncertainties on the RV in Table~\ref{targetproperties} are the standard deviations of the fitted Gaussians. The spectra from each chip were treated individually. Deriving the RV also served as an independent check of cluster membership, as all stars in the cluster are expected to have roughly the same RV. We find a mean RV of $-19.50\pm5.29$\kms, consistent with the results from \citet{alves-brito} and \citet{koch}.

\begin{table*}%
\centering
\caption{Properties of the observed targets. Boldface numbers indicate polluted targets. The uncertainties on the magnitudes are at the 3\% level.}
\begin{tabular}{rccccccr}
\hline
 ID & $\alpha$(J2000) & $\delta$(J2000) & V  &   B &   I &   U  & V$_{rad}$ \\
\hline\hline
 \textbf{1062} & 00:24:04.51 & $-$71:57:11.14 & 12.0 & 13.5 & 10.4 & 15.4 & $-22.15\pm0.24$ \\
 4794 & 00:26:28.48 & $-$71:50:12.86 & 12.4 & 13.8 & 11.0 & 15.3 & $-28.04\pm0.19$ \\
 \textbf{5265} & 00:26:57.45 & $-$71:49:13.95 & 12.1 & 13.6 & 10.6 & 15.4 & $-18.73\pm0.25$ \\
 5968 & 00:26:27.21 & $-$71:47:44.11 & 12.1 & 13.8 & 10.6 & 15.2 & $-21.76\pm0.21$ \\
 6798 & 00:23:12.87 & $-$72:10:19.00 & 12.3 & 13.7 & 10.9 & 15.2 & $-18.44\pm0.23$ \\
10237 & 00:24:45.81 & $-$72:09:10.42 & 12.5 & 13.7 & 11.2 & 14.9 & $-25.82\pm0.20$ \\
13396 & 00:22:06.69 & $-$72:07:15.77 & 12.6 & 13.8 & 11.2 & 15.0 & $-9.78\pm0.25$ \\
20885 & 00:23:05.17 & $-$72:04:27.98 & 12.6 & 13.8 & 11.3 & 15.0 & $-15.17\pm0.27$ \\
\textbf{27678} & 00:22:30.27 & $-$72:01:41.07 & 12.0 & 13.6 & 10.4 & 15.4 & $-24.86\pm0.21$ \\
\textbf{28956} & 00:23:22.79 & $-$72:01:03.83 & 11.9 & 13.4 & 10.3 & 15.3 & $-17.72\pm0.26$ \\
29861 & 00:22:07.16 & $-$72:00:32.90 & 12.6 & 13.9 & 11.3 & 15.2 & $-11.65\pm0.35$ \\
\textbf{38916} & 00:24:15.81 & $-$72:00:41.50 & 12.5 & 13.8 & 11.1 & 15.3 & $-20.72\pm0.24$ \\
\textbf{40394} & 00:24:05.09 & $-$72:00:03.35 & 12.0 & 13.5 & 10.3 & 15.3 & $-18.67\pm0.25$ \\
\hline
\end{tabular}
\label{targetproperties}
\end{table*}

\section{Abundance analysis}

\subsection{Fundamental stellar parameters}
In order to determine the fundamental stellar parameters of our targets, we adopt the traditional spectroscopic approach of combining 1D LTE atmospheric models with equivalent width (EW) measurements of a number of \feone\ and \fetwo\ lines, to enforce abundance equilibrium (determines \teff\ and microturbulence, \vmic), as well ionization equilibrium between \feone\ and \fetwo\ to determine \logg. 

EWs of all iron lines were measured using the {\tt splot} task in IRAF \citep{tody1,tody2}. Each line was identified by comparing with lines present in the synthetic spectrum that was also used for the CCF calculations. Preferably, only isolated, unblended lines were used for this purpose, and great care was taken to place the continuum at the right level. In the few cases where blends were present, the lines were appropriately deblended, using either Gaussian or Voigt profiles, depending on the strength and shape of the lines under consideration. For the first pass at parameter determination, we selected lines by comparing synthetic lines to a high-resolution spectrum of Arcturus, choosing only lines that were reproduced well in the synthesis, to minimize the impact of low-quality atomic line data. This selection of lines resulted in between 60 and 80 iron lines for each star, but since stellar parameters determined from iron lines are highly sensitive to the specific choice of lines, we chose to include only lines that were in common for at least 8 stars from the initial pass at EW measurements. In cases where lines were missing from this master linelist, great effort was put into recovering the missing lines, although this was not always possible, usually due to atmospheric emission lines or noise spikes being present in the feature. Typically 45 \feone\ and 12 \fetwo\ lines were used in each star. A cut of the full line list is presented in Table~\ref{fe-lines}. The complete list is made available online.

\begin{table}%
\centering
\caption{Cut of the full line list for the elements without reported HFS. XX.0 refers to neutral species and XX.1 to the first ionization stage. All wavelengths are given in \AA. The full table is available in the online journal.}
\begin{tabular}{rrrr}
\hline
Wavelength & Ion & log(gf) & E$_{low}$ \\
\hline\hline
\input{non-hfs-lines-cut.tex}
\end{tabular}
\label{fe-lines}
\end{table}
 
Having measured the EWs, the results were passed to the spectral analysis code MOOG (2013 version, \citealt{sneden1,sobeck,sneden2}), where interpolated, $\alpha$-enhanced ATLAS9 1D LTE models (\citealt{castelli}, [$\alpha$/Fe$]=+0.4$ dex) were used in an initial pass at determining the stellar parameters. For each star, we determined \teff\ by requiring no correlation between the abundance of \feone\ and lower excitation potential, E$_{low}$. As a starting point we used the mean value of the photometric \teff, using the calibrations of \citet{ramirez} and \citet{gonzales-hernandez}. The photometric \teff\ was derived assuming an interstellar reddening of E(B$-$V) $ = 0.04$ (see \citealt{grundahl} and references therein), where we used the calibration of \citet{taylor} to convert from E(B$-$V) to E(V$-$I). The photometric \teff\ is quoted in Table~\ref{photteff} for reference. \vmic\ was determined by requiring no correlation between the abundance of \feone\ and reduced equivalent width, log(EW/$\lambda$). Finally, the surface gravity was determined by enforcing ionization equilibrium between \feone\ and \fetwo. We used atomic data from version 4 of the linelist for the Gaia-ESO survey (GES, Heiter et al. in prep.), with the exception of oscillator strengths for \fetwo, which were adopted from the work of \citet{melendezgf}, in cases where the GES linelist used other sources. Using \loggf\ values from a single study were found to yield more homogeneous \fetwo\ abundances. After the initial best-fitting model had been determined, we proceeded to measure the abundances of O, Na, Mg, Al, Si, Cr, Ti and Ni for each star (see Sect.~\ref{sec:elem}). 

\begin{table}%
\centering
\caption{Photometric \teff's for each sample star. Subscript "R" and "G" refer to the calibrations of \citet{ramirez} and \citet{gonzales-hernandez}, respectively.}
\begin{tabular}{lccc}
\hline
ID 	& $T_{\rm eff,B-V,R}$ & $T_{\rm eff,V-I,R}$ & $T_{\rm eff,B-V,G}$ \\
\hline\hline 
      \textbf{1062}   &   4036  &    4098  &    4131 \\
      4794   &   4193  &    4228  &    4245 \\
      \textbf{5265}   &   4033  &    4096  &    4129 \\
      5968   &   3746  &    3899  &    3952 \\
      6798   &   4151  &    4192  &    4214 \\
      10237  &   4347  &    4371  &    4366 \\
      13396  &   4305  &    4331  &    4332 \\
      20885  &   4342  &    4366  &    4362 \\
      \textbf{27678}  &   3950  &    4033  &    4074 \\
      \textbf{28956}  &   3983  &    4057  &    4096 \\
      29861  &   4278  &    4305  &    4310 \\
      \textbf{38916}  &   4236  &    4266  &    4277 \\
      \textbf{40394}  &   3987  &    4060  &    4098 \\
\hline
\end{tabular}
\label{photteff}
\end{table}

Since this work is concerned with cool giants with a variation in light elements, it is desirable to take this into account in the atmospheric modelling, as these elements are important electron donors. As the strength of many lines are sensitive to the electron pressure in the atmosphere, it is important that the number of free electrons are properly accounted for. Failing to do so can have significant impact on the derived abundances which may, in turn, lead to erroneous conclusions, as illustrated \eg\ by \citet{mucciarelli} in the case of NGC2419. In particular, the correct abundance of Mg is important, as this is the most important electron donor.  

Therefore, with an initial set of abundances determined from the use of ATLAS9 models, we computed new, tailored atmospheric models for each star with the abundance pattern and fundamental parameters derived as described above. We utilized the GNU Linux-ported version of the ATLAS12 code \citep{kurucz,atlas1,atlas2} for this purpose, where the main difference between this code and ATLAS9 is that the latter uses a precomputed set of opacity distribution functions, calculated for a fixed abundance mixture, whereas ATLAS12 uses opacity sampling, which allows for the opacities to be calculated on-the-fly for any given element mixture. The models were subsequently ported to a MOOG-friendly format and the stellar parameters were re-derived. Since the atmospheric structure changes compared to the initial models, a few iterations were typically required to re-establish ionization and abundance equilibrium with the ATLAS12 models. In each iteration, the entire set of measured abundances was updated, to ensure as close a match to the free electron density as possible. 

\begin{figure}
\centering
\begin{tabular}{c}
\includegraphics[width=\columnwidth,trim= 2cm 13cm 1.5cm 3cm, clip=true]{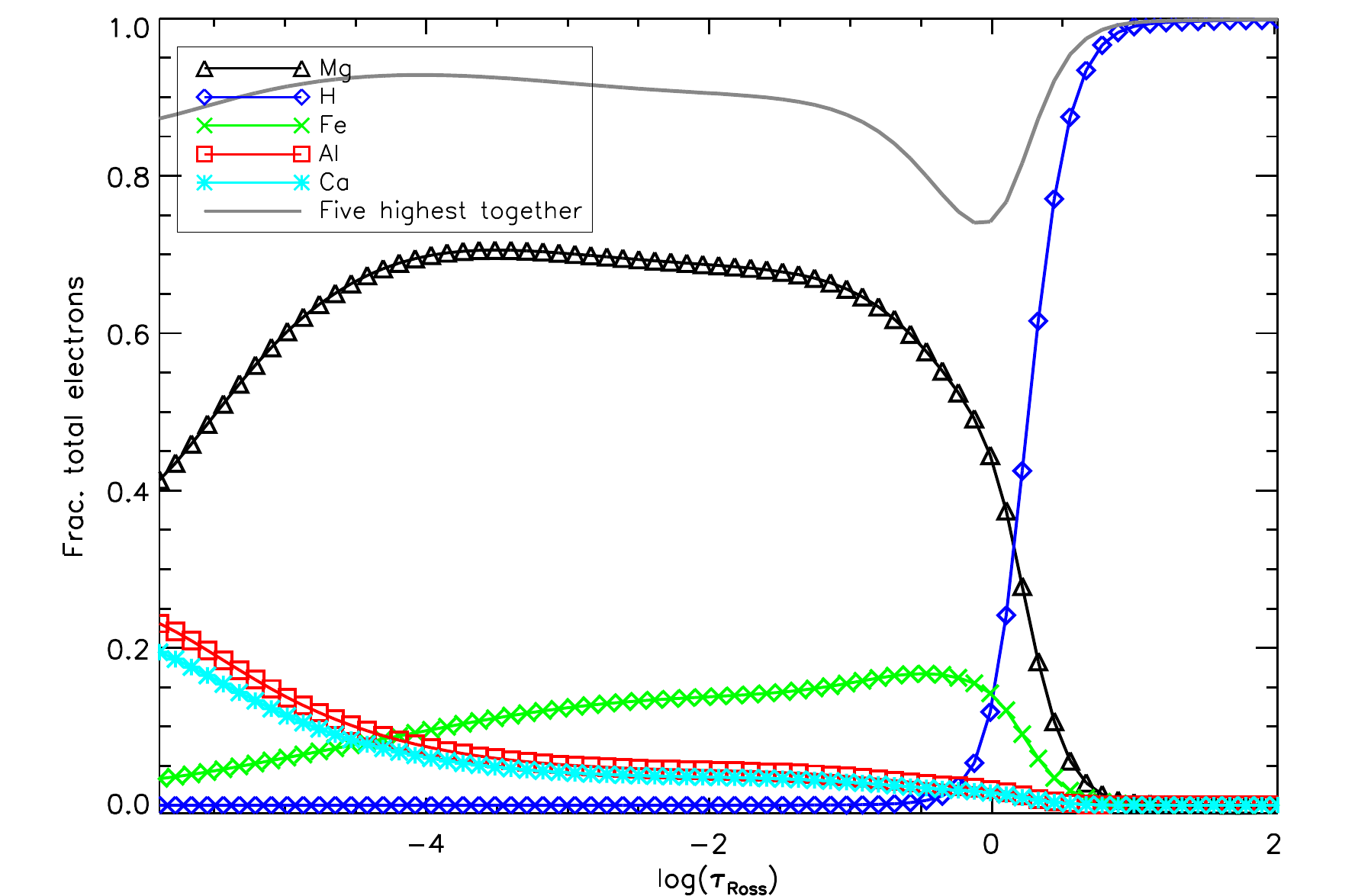}\\
\includegraphics[width=\columnwidth,trim= 2cm 13cm 1.5cm 3cm, clip=true]{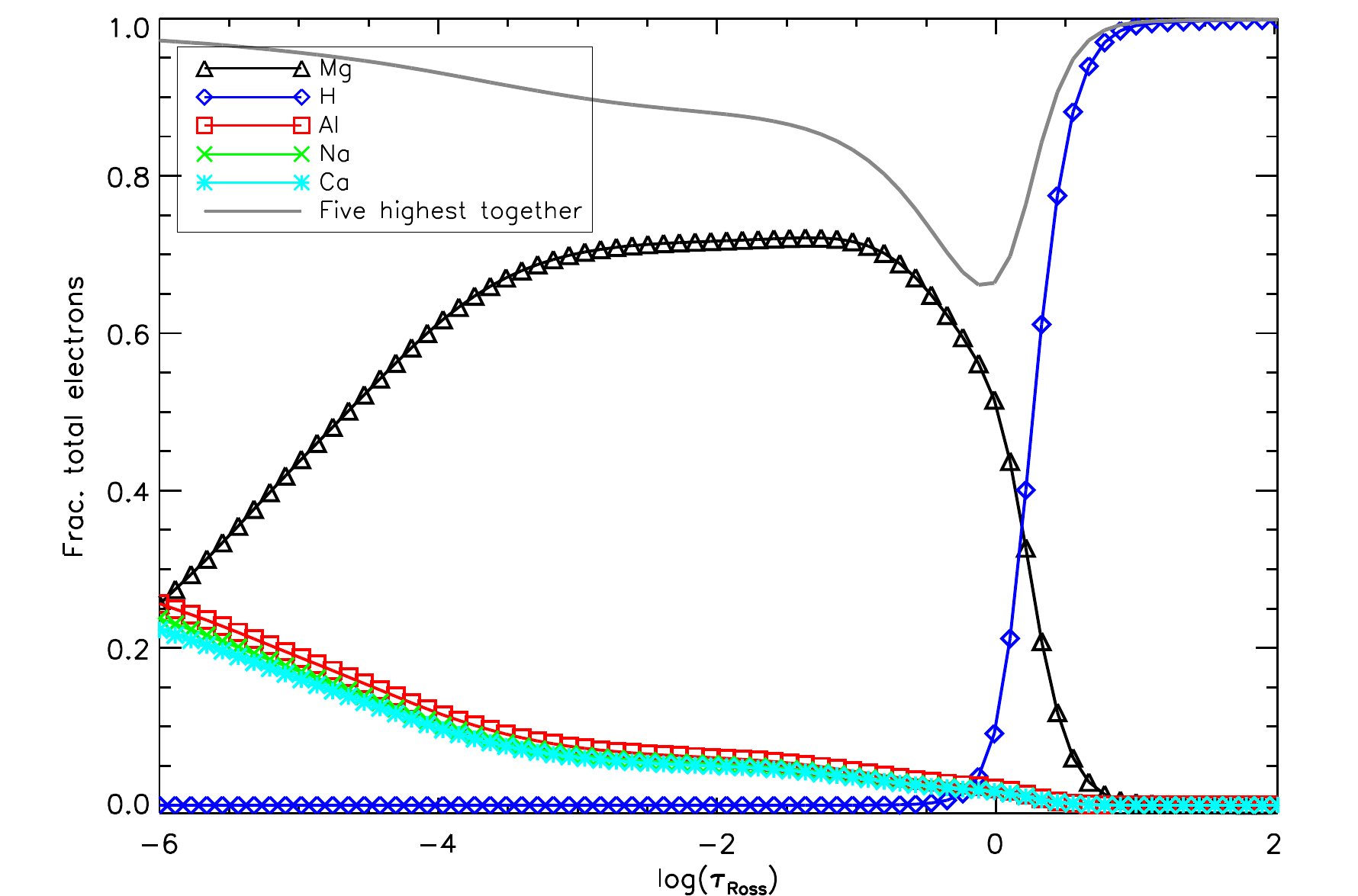}%
\end{tabular}
\caption{The fractional contribution of free electrons for the top five electron donors as a function of optical depth. The solid, gray line shows the total contribution from the 5 species. Top: 10237, which is the most Mg-depleted star. Bottom: 38916, the most Mg-enhanced star.}%
\label{mg-electron}%
\end{figure}

\begin{figure}%
\centering
\includegraphics[width=\columnwidth,trim= 2cm 13cm 1.5cm 3cm, clip=true]{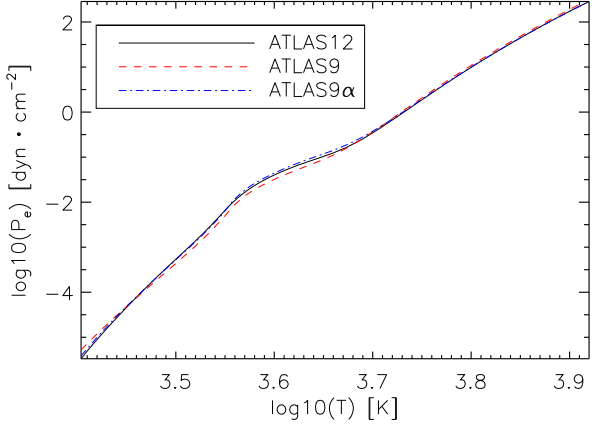}%
\caption{Electron pressure as a function of temperature for three different models of the star 10237. The solid, black line shows the ATLAS12 model with tailored abundance pattern, and the red, dashed line shows the interpolated, scaled solar ATLAS9 model. Finally, the $\alpha$-enhanced ATLAS9 model is shown as the blue dot-dashed line.}%
\label{at12vsat9}%
\end{figure}

The impact of wrong abundances of the main electron donors is lower in a high-metallicity case like 47 Tucanae, since the variation in Mg is small, compared to the case of NGC2419, where the variation in [Mg/Fe] spans almost 2\,dex, but the variation is still visible. Inspecting Fig.~\ref{mg-electron}, where we plot the fractional contributions to the total number of free electrons for the five most significant electron donors as a function of optical depth, even the effect of small variations in Mg to the electron budget can be seen. The two panels show star 10237 (top), which is the most Mg-depleted star, [Mg/Fe]\,=\,0.32\,dex and star 38916 (bottom), the most Mg-enhanced star, [Mg/Fe]\,=\,0.52\,dex. Slight variations in the electron pressure can also be seen, compared to using a standard scaled solar ATLAS9 model as can be seen in Fig.~\ref{at12vsat9}. On the other hand, the ATLAS12 model can hardly be distinguished from the equivalent ATLAS9 $\alpha$-enhanced model.

For the less evolved giants in our work \ie, \teff\,$>4000$\,K, \logg\,$>1.0$\,dex, we found that the differences between using interpolated ATLAS9 $\alpha$-enhanced models and dedicated ATLAS12 models were negligible. However, slight differences in the electron pressure were found for the low gravity models. These differences had a small effect on the derived \logg\  values, where we found differences of up to 0.13\,dex. This may, in turn, affect the derived abundances slightly, especially for pressure-sensitive species. However, the variation in \logg\ is smaller than our typical uncertainty. The effect on temperature and \vmic\ were found to be negligible in all cases. 

\begin{figure}%
\centering
\includegraphics[width=\columnwidth,trim= 2cm 13cm 1.5cm 3cm, clip=true]{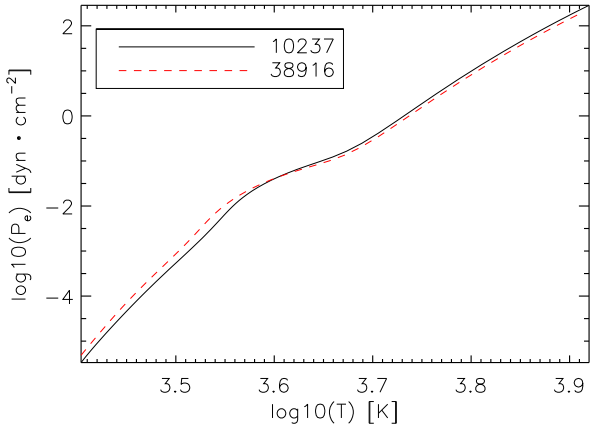}
\caption{Difference in electron pressure as a function of temperature between star 10237 and star 38916.}%
\label{at12vsat12}%
\end{figure}

\subsubsection{Parameter uncertainties}
To estimate the uncertainties on the fundamental parameters of our stars, we adopted the uncertainty of the fitted slope of $\log\epsilon({\rm Fe})$ vs. EW and EP and performed the following exercise.

For a representative star (6798) we perturbed the best fitting model by $\pm200$\,K in steps of 50\,K for \teff, and $\pm0.3$\,\kms\ in steps of 0.1\,\kms\ for \vmic, around the adopted best values. Only one of the parameters was perturbed at a time. Each of the perturbations will change the slope of the best-fitting line. This exercise allows us to produce the $\Delta$\teff\ vs. EP-slope and the $\Delta$\vmic\ vs. EW-slope relations by making a linear fit to the observed changes. From this, we can easily calculate $\Delta$\teff\ and $\Delta$\vmic\ corresponding to the uncertainty of the zero-slope model. This corresponds to uncertainties of 80\,K in \teff\ and 0.1\,\kms\ in \vmic, which we adopted for all stars, after checking that these values did not depend on the absolute values of the stellar parameters. To check if an offset of the metallicity scale affected \teff\ and \vmic, we also perturbed the overall metallicity, [M/H], of the star by changing it by $\pm$0.15 dex, which is the typical line-to-line scatter of our iron abundances, which were use as a proxy for overall metallicity for this purpose. As can be seen in Table~\ref{sysuncert}, this had only a negligible effect on the remaining parameters.

To determine the uncertainty in \logg\ we adopted a similar approach, but this time inspecting the changes of the abundance of \fetwo, as it is strongly influenced by the surface gravity. Again, we perturbed the models by $\pm0.3$\,dex in steps of $0.1$ and used the change in iron abundance vs. change in \logg\ to estimate the uncertainty. The change in \logg\ that corresponded to a difference between \feone\ and \fetwo\ equal to the standard deviation of the best fitting \fetwo\ abundance was taken as the uncertainty of \logg, resulting in a value of 0.20\,dex.

Whenever one of the model parameters is perturbed, it will influence the remaining parameters to some extent. For instance, changing \teff\ will also introduce small changes to the abundance vs. EW slope as well as to the derived mean iron abundance. To give an indication of the sensitivity of each parameters to changes in any of the others, we give in Table~\ref{sysuncert} the changes to each of the fundamental parameters when one is changed within the uncertainties, while the others are kept fixed. The changes are given in the sense $(best~fit - perturbed)$. The quoted $\Delta$\logg\ reflects the change needed in \logg\ to bring the \fetwo\ abundance back into agreement with \feone, within the accepted difference between the two species (0.04\,dex). As can be seen, the small parameter changes affect each other; most noticeably, changes in \teff\ have a significant influence on \logg, as \teff\ changes have a strong impact on the \fetwo\ abundance. Thus, the \logg\ needs to be changed by an appreciable amount to reestablish ionization equilibrium. On the other hand, \feone, \vmic, and \teff\ itself are hardly changed from perturbations of the other parameters. We adopt the following values as the final uncertainties of our fundamental parameters, $\sigma_{T_{{\rm eff}}}=\pm80$\,K, $\sigma_{\log g}=\pm0.20$\,dex and $\sigma_{\xi_t}=\pm0.10$\,\kms. 

We note that the \fetwo\ lines appear to be affected slightly more by temperature changes, compared to changes in \logg. We investigated this further, using the \texttt{MyGisFOS} analysis software \citep{mygisfos}, which produced similar results. In addition, the changes in the \fetwo\ abundances with changing parameters are not associated with an increased line-to-line scatter, but merely reflects a change in the mean abundance.

NLTE effects on iron may potentially influence the derived stellar parameters, as illustrated by e.g. \citet{bergemann} and \citet{lind2}. In order to check whether this was important for our stars, we used the \texttt{INSPECT} online database\footnote{\url{http://www.inspect-stars.net}}, which requires input stellar parameters as well as the EW of the line under investigation. The database then interpolates in a large grid of models to provide a NLTE correction. The corrections for iron have been computed based on the work of \citet{bergemann} and \citet{lind2}. 
We chose the star 10237 as a representative example for this exercise. Of the 42(12) FeI(FeII) lines used for the parameter determination, only 6 FeI and 12 FeII lines were present in the database. However, as the NLTE corrections will be very similar for different lines with the same EP and \loggf, we adopted corrections for lines which matched a line with computed NLTE corrections, within 0.2 eV in EP and 0.2 in \loggf, to assess the magnitude of NLTE on the parameters. Nevertheless, 16 of the lines used in the LTE analysis had to be discarded, as no satisfactory matches could be found. The NLTE corrections were of the order of 0.02\,dex. Applying these corrections to the matching lines resulted in parameter changes of $\Delta$\teff\,$=45$\,K, $\Delta$\logg\,$=0.05$\,dex, $\Delta$\vmic\,$=0.02$\,\kms\ and $\Delta$[Fe/H]\,$= 0.05$\,dex, which should be compared to typical uncertainties of 80\,K, 0.2\,dex, 0.1\,\kms\ and 0.15\,dex, respectively. We thus do not consider NLTE effects on neither \feone\ nor \fetwo\ to have a significant effect on our derived stellar parameters. 

\begin{table*}%
\centering
\caption{Changes to the fundamental parameters when perturbing a typical star (6798) by the estimated uncertainties. Changes given in the sense $(best~fit-perturbed)$.}
\begin{tabular}{lccccc}
\hline
Change/result & $\Delta$\teff & $\Delta$\vmic & $\Delta$\logg & $\Delta$\feone & $\Delta$\fetwo \\
\hline\hline
$\Delta$\teff $=\pm80$K & - & $+0.02/-0.03$ & $-0.20/+0.20$ & $+0.00/-0.02$ & $+0.16/-0.18$ \\
$\Delta$\vmic $=\pm0.1$\kms & $+34/-30$ & - & $-0.05/+0.05$ & $+0.02/-0.04$ & $+0.03/-0.03$ \\
$\Delta$\logg $=\pm0.20$ dex & $-33/+38$ & $-0.02/+0.02$ & - & $-0.04/+0.03$ & $-0.12/+0.12$ \\
$\Delta$[M/H] $=\pm0.15$ dex & $\pm8$ & $\pm0.01$ & 0.00 & $-0.03/+0.03$ & $-0.06/+0.07$ \\
\hline
\end{tabular}
\label{sysuncert}
\end{table*}

\subsection{Element abundances}
\label{sec:elem}
In the initial pass on abundance measurements, using the ATLAS9 models, we derived abundances of the $\alpha$-elements (O, Mg, Si, Ca, Ti) as well as Na, Al, Cr, Fe, Ni and Zn. With the exception of O and Na, all initial abundances were derived using EW measurements, as for the iron lines. After these abundances had been determined from the ATLAS9 models, tailored ATLAS12 models were calculated, matching the measured abundance pattern as described above. All abundances were subsequently re-derived in the fashion described in the following.

\subsubsection{Equivalent width measurements}
The abundances of Si, Ca, Ti, Cr, Fe, Ni and Ce were derived with MOOG using EW measurements. With a few exceptions, all atomic line data were adopted from version 4 of the GES linelist (Heiter et al., in prep.). For Ti we used the recent results of \citealt{lawler} (Ti\,{\sc I}) and \citealt{wood} (Ti\,{\sc II}). As with the iron lines, de-blending was taken into account when needed, but as far as possible only clean lines were used. Whenever individual lines showed large deviations from the mean abundance, they were inspected in detail by comparing with a synthetic spectrum as well as with an observed spectrum of a star with similar fundamental parameters, in which the line was not discrepant. In most cases, this resulted in a re-measurement of the discrepant line, which would remedy the deviation. These deviations were in most cases due to improper continuum placement. The star-to-star comparison was particularly useful here, as it allowed to determine the continuum more easily in cases where one spectrum suffered from atmospheric emission in the continuum area near the line under investigation. In some cases, this approach also enabled us to identify overlooked blends with atmospheric lines or noise spikes in the lines themselves. In these cases, the lines were discarded.

For Ce, only one line at 5274.23\,\AA\ was used to derive the abundance. If it is not deblended properly or continuum placement is not being done carefully, the derived abundance may be systematically off. However, checking against a spectral synthesis, it was found that the abundances from EW measurements agreed with the abundances derived from syntheses to within 0.1\,dex. Thus, we trust that the de-blending was done in a reliable way. Based on the synthesis deviation, we adopt 0.1\,dex as the measurement uncertainty of the abundance for this element.	

\subsubsection{Spectral synthesis}
Not all element abundances can be derived reliably from EW measurements due to line-blending issues if, for instance, these are strong, or, if the lines under investigation are in regions with a large number of weak molecular features. In order to derive reliable abundances one has to employ spectral synthesis. This technique was used to derive abundances of O, Na, Mg, Zn, Mo, Ru and Dy.

The derivation of the oxygen abundance is difficult due to the very few lines of appreciable strength in the optical range. In our case we utilized the two forbidden [OI] lines at 6300.3\,\AA\ and 6363.8\,\AA. The former suffers from a close blend with a Ni line, whereas the latter is in the extended wings of the Ca autoionization line at 6360\,\AA. Both things make it difficult to use EW measurements and deblending methods to reliably derive the abundances. MOOG was used to synthesize both lines. For the Ni\,I line at 6300.34\,\AA, blending with the [O] 6300\,\AA\ line, we used the data from \citet{johansson}. For each star we adopted the Ni abundance derived from the EW measurements, updating after each iteration. Initially, we synthesized a 20\,\AA\ window around the line of interest. This was done to ensure an accurate continuum placement. With the continuum set, the O abundance was adjusted until a minimum of the $\chi^2$ value between the observed and synthetic spectrum was found. When synthesizing the lines we also allowed for small variations in the macroturbulent broadening to ensure the best possible fit to the observed line profile, but in most cases it was found that a single value could be applied to both lines. For slow rotators such as evolved red giants, we cannot reliably disentangle rotational broadening and macroturbulence and we treat them as a single broadening, which we will refer to as macroturbulent broadening throughout the paper. The presence of a telluric absorption line at 6300.598\,\AA\ could in principle be contaminating the oxygen line, but the radial velocity of the stars were sufficient to shift the oxygen away from the contaminated region.

\begin{figure*}%
\centering
\includegraphics[width=\textwidth,trim= 2cm 12.5cm 0cm 6cm, clip=true]{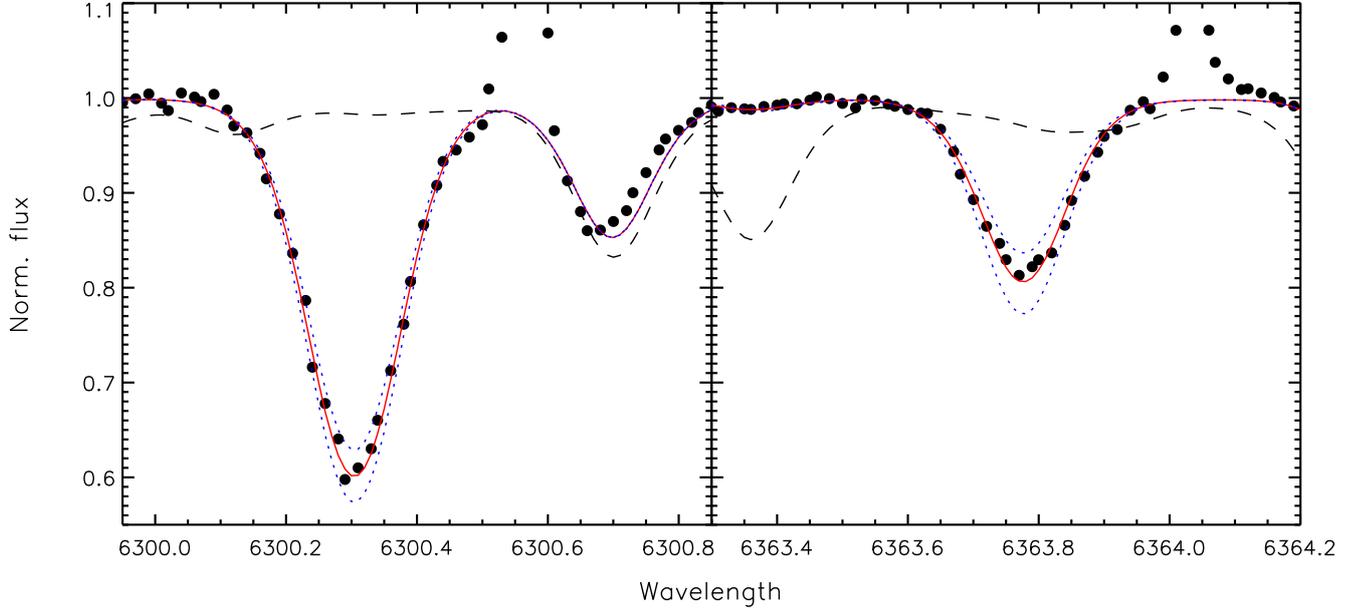}%
\caption{Example of the oxygen synthesis for the two lines used to derive the O abundance in this work. Here shown for star 10237. Solid red line shows the optimal fit and the dotted blue lines changing the O abundance by $\pm0.07$ dex. The black, dashed line show a synthesis without oxygen.}%
\label{osynth}%
\end{figure*}

To assess the uncertainty of the measurement, the abundance was subsequently changed until a significant deviation from the best fit was observed, as judged by eye. This typically required changes in the oxygen abundance on a level of 0.07\,dex, which can easily be distinguished from the best fitting model, as illustrated in Fig.~\ref{osynth}, where we also show a synthesis without any oxygen in the atmosphere. That some lines are seen to increase in strength when lowering the O abundance is due to less carbon being locked in CO molecules, thus increasing the strength of C$_2$, CH and CN molecular lines also present in this region. Also, sky emission lines are visible redwards of each of the two oxygen lines. These did, however, not affect our derived abundances.

A discrepancy of the abundance between the 6300\,\AA\ and 6363\,\AA\ lines has been observed in the Sun, as well as other dwarf stars, but in the case of giants, both lines give consistent abundances (see \citealt{caffau}). In studies of the Sun, \citet{caffau3} found that NLTE effects were negligible for these lines at solar metallicities and low \teff\ (see also \citealt{kiselman}). This result is expected to hold also for the parameter range of this sample of stars \citep{fabbian}. In the case of the star 13396 only the 6300\,\AA\ line was used, due to atmospheric emission being present in the 6363\,\AA\ line. In two other cases (stars 5265 and 6798), some emission was present in the wings of both oxygen lines, so the derived abundance should be considered as a lower limit. For stars 20885 and 38916, strong emission was present in both lines, which made it impossible to derive a reliable oxygen abundance. 

\begin{figure}%
\centering
\includegraphics[scale=1.4,trim= 2cm 12.5cm 11cm 4cm, clip=true]{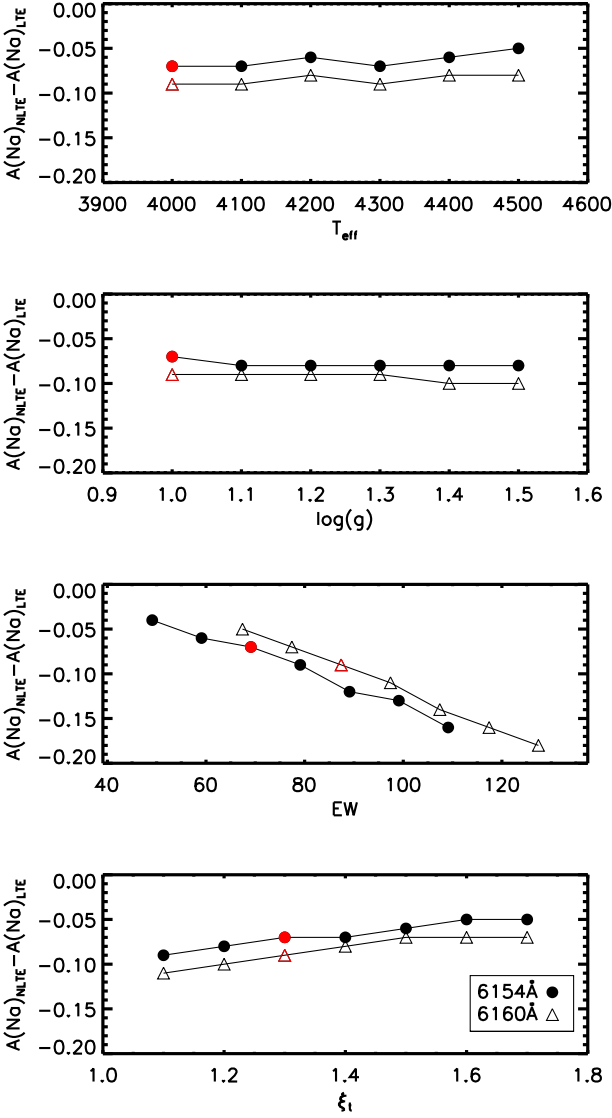}%
\caption{NLTE corrections (A(Na)$_{{\rm NLTE}}$-A(Na)$_{{\rm LTE}}$) as a function of \teff, \logg, EW and \vmic. The red points shows the actual values for the star 6798.}%
\label{na-nlte}%
\end{figure}

The synthesis of the sodium lines followed a similar approach. The fact that the stars in this work are fairly evolved giants, combined with the Na lines typically being strong ($>$70m\AA), makes them susceptible to non-LTE effects. To assess this, we  again used the {\tt INSPECT} database. For the EW we used the measured width of the best-fitting, \emph{synthetic} line, calculated as a single line, without including any lines in the vicinity to by-pass blending issues. This provides a NLTE correction for each line under investigation, using the computations from \citet{lind}. Following the recommendation in the mentioned study, we only used the 6154\,\AA\ and 6160\,\AA\ lines. However, the lowest \teff\ and \logg\ values available in the grid (4000\,K, 1.00\,dex), are somewhat higher than the parameters of the most evolved giants treated in our work. For these stars, we used the 4000\,K, 1.00\,dex NLTE correction, which in the most extreme cases are off by 150\,K and 0.55\,dex respectively. In Fig.~\ref{na-nlte} we plot the NLTE corrections vs. different stellar parameters as well as the EW for the two lines used. We began at the extremes of the parameters available in the precomputed NLTE grid and used the EWs from star 6798, as this object is representative of our sample and has the lowest \teff\ that still falls within the grid calculated by Lind et al. The value of \logg\ of 0.9\,dex is marginally outside the available parameters, but, as can be seen in Fig.~\ref{na-nlte}, the size of the correction is mostly sensitive to the line EW, and not the actual stellar parameters within the range in our sample. In each of the panels only one parameter is changed, whereas the remaining are kept fixed at the values shown in red in the remaining panels. The variation in the NLTE corrections due to changes in \vmic\ are also non-negligible, but at least a factor of two smaller than the corrections from varying the EW. We note that in no case does our derived value of \vmic\ extend beyond the available grid. Thus, we are confident that using the corrections from the lowest possible grid values will not deviate significantly from the true corrections, as they are insensitive to \teff\ and \logg\ changes. The size of the corrections was never more than $-0.15$\,dex.

For the final abundance determination of Mg we performed spectral synthesis of the lines at 6318.72\,\AA, 6319.24\,\AA, and 6319.50\,\AA, using the line data of \citet{melendez}. These lines are in the vicinity of the atmospheric lines at 6318.85\AA\ and 6319.59\AA. As for the oxygen lines, we checked whether the radial velocity of each star was high enough to shift the Mg lines away from the tellurics. In all stars except 29861, at least two of the three Mg lines were free of telluric contamination and reliable Mg abundances could be derived. For 29861	only the 6319.24\,\AA\ line was free of contamination. For this line we adopt the mean uncertainty derived from the remaining lines. Using the \texttt{MULTI} code \citep{carlsson,korotin} we checked for the impact of NLTE on the used lines by performing an NLTE synthesis for the two stars with the most extreme parameters. We found that the corrections were $\leq0.05$ dex, which we considered negligible compared to our typical uncertainties and thus only a standard LTE synthesis was applied.

The abundances of Zn, Mo, Ru and Dy were also derived from a standard spectral synthesis. The Zn line at 4810.53\,\AA\ is strongly blended with a Cr\,I feature and the Dy line at 4890.10\,\AA\ is suffering from a blend with a Ti\,I line as well as several weak C$_2$ and CN molecular lines making a standard EW analysis overestimate the abundances. Similar blending issues are affecting the Mo and Ru lines. The uncertainties on the abundances includes the uncertainty in the continuum placement (less than 0.1\,dex variation) as well as the change in abundance needed to produce a noticeable deviation from the best-fitting abundance, as judged by eye.

\subsubsection{Ba and Al NLTE synthesis}

\input{nlte.tex}

\begin{figure*}%
\centering
\includegraphics[width=\textwidth,trim= 2cm 12.5cm 0cm 6cm, clip=true]{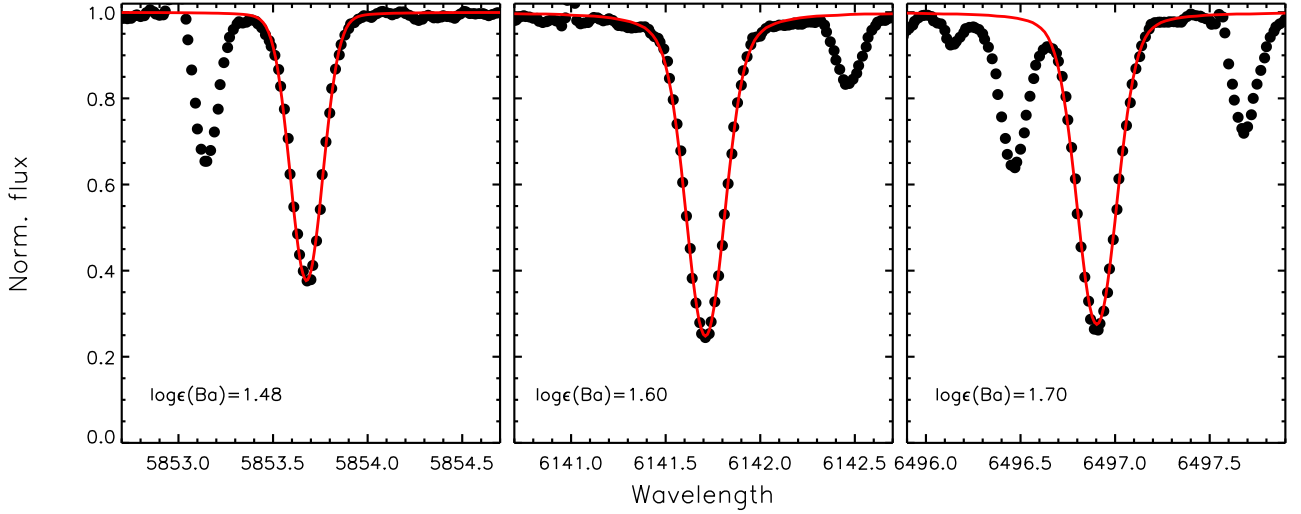}%
\caption{Example of the NLTE line profile fitting of the three Ba lines used here. Shown is the star 10237. In each panel, we give the abundance used in the fit.}%
\label{basynth}%
\end{figure*}

\begin{figure}%
\centering
\includegraphics[width=\columnwidth,trim= 2cm 13cm 1cm 3cm, clip=true]{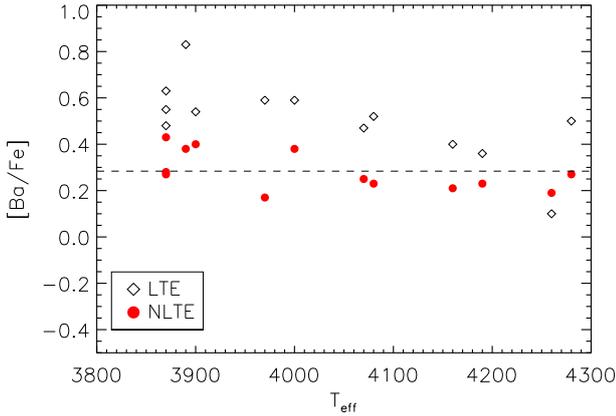}%
\caption{[Ba/Fe] vs. \teff\ for both LTE (black) and NLTE (red). Taking NLTE into account removes the correlation between abundance and \teff.}%
\label{bacorr}%
\end{figure}

\begin{figure}%
\includegraphics[trim= 2.2cm 12.7cm 4.5cm 8.7cm, clip=true]{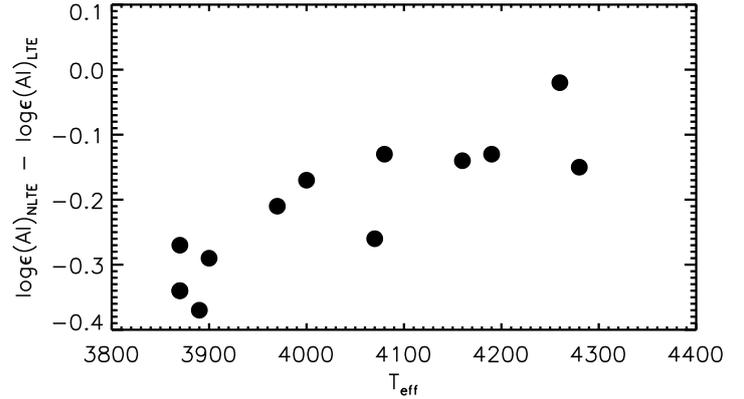}%
\caption{Size of the NLTE correction for Al vs. \teff. }%
\label{al-nlte-vs-teff}%
\end{figure}

\subsubsection{HFS lines}
Lines of a number of elements of interest (Sc, V, Mn, Co, Cu, Y, Zr, La, Pr, Nd, Eu) suffer from HFS, making a standard EW analysis overestimate the actual abundance. These lines were analyzed by calculating a grid of syntheses around each line of interest, using the {\tt SYNTHE} code. For each line, syntheses were computed, varying the abundance of the element by $\pm0.5$\,dex in steps of 0.05\,dex around the starting abundance, taken to be the solar-scaled value for the metallicity of the star. Each synthesis was subsequently convolved with a Gaussian (details to follow) to match the observed spectra. HFS data were taken from the version 4 of the GES line list with the exception of Y and Zr, which was taken from \citet{mcwilliam2013}. A cut of the full line list, split in HFS-components, is given in Table~\ref{hfs-elements}. 

\begin{table}%
\centering
\caption{Cut of the linelist of HFS-elements measured. XX.0 and XX.1 indicates neutral and ionized species respectively. The full table is available online. With the exception of Barium, all lines have been split into individual HFS components in this list. All wavelengths given in \AA.}
\begin{tabular}{ccrr}
\hline
Wavelength & Ion & log(gf) & E$_{low}$ \\
\hline\hline
\input{hfs-lines-sorted-cut.tex}
\end{tabular}
\label{hfs-elements}
\end{table} 

For each star, fitting regions were determined for each line, as well as carefully selected continuum intervals bracketing each feature of interest. Using the continuum regions, the observed spectra were pseudo-normalized, before determining the best fitting abundance by minimization using the \texttt{MINUIT} package \citep{minuit1,minuit2}. The fitting routine (\texttt{FITPROFILE}) allowed for small velocity shifts of the lines ($\pm3.0$ \kms) as well as small adjustments of the continuum level to ensure an optimal fit when performing the minimization. \texttt{FITPROFILE} will be described in more detail in Paper II, as it was developed for fitting the MgH molecular bands.

When performing the fitting, our software provides for multi-parameter fits, allowing the velocity broadening of the lines to be included as a fitting parameter. This assumes a single, Gaussian convolution, taking into account instrumental broadening as well as rotational velocity and macroturbulence as a whole. Initially, the macroturbulent broadening had been determined by synthesizing five clean \feone\ lines across the wavelength coverage of the spectrograph. This exercise already yielded broadenings with a significant line-to-line scatter. Also, by inspecting the individual fits for the various HFS lines, it was noticed that the average value of the macroturbulent broadening determined from the \feone\ lines was underestimating the broadening needed to produce satisfactory fits to the line profiles. Thus, we decided to also include the macroturbulent broadening as a fitting parameter, allowing for a maximum variation of $\pm1.5$\kms\ around the mean value initially determined.

Adjusting the broadening for individual lines has the potential of affecting the abundances in a non-negligible fashion. We therefore tested the behavior of the abundances, keeping the broadening fixed at the mean value, as well as when leaving it free. The mean value of the abundances changed by at most 0.05 dex, well below the typical line-to-line scatter, but when the broadening was left free, the scatter became significantly lower. Further, the mean values of the fitted macroturbulence did not show a large scatter, neither across the entire set of lines nor when inspected on a element-by-element basis, the typical standard deviation being on the order of 0.6 \kms. Also, as suspected from the initial determination of the macroturbulent broadening, the fitted value were found to be larger.

In addition, studies of 3D hydrodynamical model atmospheres confirm that the velocity fields differ at different depths in the atmosphere \citep{asplundVel,collet}, which is not captured in 1D atmospheres when adopting a single value for the macroturbulence. Thus, it we consider it justified to allow for a (modest) variation of the macroturbulent broadening when fitting spectra with 1D models. The final mean values of the macroturbulent broadening are given in Table~\ref{fundam}.

\subsubsection{Abundance uncertainties}

In order to derive reliable conclusions from the abundance pattern of the cluster stars, a realistic treatment of the uncertainties is required. We took the following approach.

For each element, we derived the mean value of the abundance and assumed the random uncertainty of this value as the standard error of the mean, $\sigma_{{\rm rand}}=\sigma/\sqrt{N}$, with $\sigma$ being the standard deviation of our measurements, and $N$ the number of lines used.

The random uncertainties of the fundamental parameters will also affect the derived abundances, so these do also need to be taken into account properly. Thus, we re-derived the abundances using atmospheric models perturbed by the aforementioned parameter uncertainties. We then calculated the change in abundance on the [X/Fe] scale, relative to our best-fitting model, using the \citet{asplund} solar abundances. We use the abundance of \feone\ for neutral species and \fetwo\ for ionized species. For each parameter perturbation we then adopted the difference from the best-fitting mean value as the uncertainty of the abundance caused by that particular parameter. The uncertainties of the abundance ratios were then calculated as:

\begin{equation}
\sigma_{\rm{param, [X/Fe]}} = \sqrt{\sigma_{\Delta T_{\rm eff}}^2 + \sigma_{\Delta\xi_t}^2 + \sigma_{\Delta\log g}^2 + \sigma_{\Delta{\rm [M/H]}}^2}
\label{eq:uncert}
\end{equation}

This was done for all elements. Since our stars are very close to each other in fundamental parameters, we only performed this exercise for star 6798 as it falls in the middle of the range considered. The exception is Ce, where we do not have measurements from this star. Instead, we adopted the result from star 10237 for this particular element. The deviations are quoted in Table~\ref{abuuncert}. We note that this exercise was also performed with full NLTE analysis for Al and Ba. The influence of the parameter uncertainties on the abundance ratios for 6798 were then assumed to be representative for our complete sample and adopted for all stars.

To yield the total uncertainty for each abundance ratio, we added the random error on [X/H] and [Fe/H] to the uncertainty from the stellar parameters, resulting in an uncertainty of

\begin{equation}
\sigma_{{\rm tot}} = \sqrt{\sigma_{\rm{param, [X/Fe]}}^2+\sigma_{{\rm rand, [X/H]}}^2 + \sigma_{{\rm rand, [Fe/H]}}^2}
\label{eq:uncert_tot}
\end{equation}

\begin{table*}%
\centering
\caption{Changes to derived abundances relative to best-fitting model, when perturbed by the uncertainties on the fundamental stellar parameters. Shown for the star 6798.}
\begin{tabular}{lrrrrrrrrr}
\hline
 & \multicolumn{2}{c}{$\Delta$\teff\ [K]} & \multicolumn{2}{c}{$\Delta$\logg\ [dex]} & \multicolumn{2}{c}{$\Delta$\vmic\ [\kms]} & \multicolumn{2}{c}{$\Delta$[M/H]} & $\sigma_{\rm {param}}$\\
Elem. & $+80$ & $-80$ & $+0.2$ & $-0.2$ & $+0.10$ & $-0.10$ & $+0.15$ & $-0.15$ & \\
\hline\hline
\input{abundance_ratio_diff_v7.tex}
\hline
\end{tabular}
\label{abuuncert}
\end{table*}

We also tested the effect of using ATLAS12 models vs. ATLAS9 models on the abundances of a few selected species. For this exercise we used star 1062 as a test-case. We report the results of this exercise in Table~\ref{abumodels}, and as might be expected, the differences between the tailored model and the $\alpha$-enhanced model are small, whereas more pronounced effects are seen if one simply uses a model atmosphere with a solar abundance mixture scaled to the metallicity of the star.

\begin{table}%
\centering
\caption{Changes to derived abundances, relative to a tailored ATLAS12 model when using ATLAS9 scaled solar and ATLAS9 with $+0.4$ dex $\alpha$-enhancement. Both models have the same fundamental parameters. Shown here for star 1062.}
\begin{tabular}{lrr}
\hline
 Elem. & ATLAS9 & ATLAS9$\alpha$ \\
\hline\hline
$\Delta$[O/Fe] & 0.05 & $-$0.02 \\
$\Delta$[Na/Fe] & $-$0.07 & 0.01 \\
$\Delta$[Mg/Fe] & $-$0.03 & 0.00 \\
$\Delta$[Si/Fe] & $-$0.02 & $-$0.02 \\
$\Delta$[Ca/Fe] & $-$0.06 & $-$0.05 \\
$\Delta$[Ti\,I/Fe] & $-$0.02 & $-$0.04 \\
$\Delta$[Ti\,II/Fe] & 0.01 & $-$0.02 \\
$\Delta$[Cr/Fe] & $-$0.05 & $-$0.02 \\
$\Delta$[Fe\,I/H] & 0.06 & $-$0.02 \\
$\Delta$[Fe\,II/H] & 0.11 & $-$0.06 \\
$\Delta$[Ni/Fe] &  0.00 & $-$0.01 \\
$\Delta$[Zn/Fe] & $-$0.05 & $-$0.04 \\
$\Delta$[Ce/Fe] & 0.02 & $-$0.02 \\
\hline
\end{tabular}
\label{abumodels}
\end{table}

\section{Results}
The fundamental parameters for our targets are presented in Table~\ref{fundam}. Our targets span a rather narrow range in all parameters, with the largest spread seen in \logg. We find a mean metallicity of \feh\ $=-0.78\pm0.07$ dex, in excellent agreement with the study of \citet{koch}, who found \feh\ $=-0.76\pm0.01\pm0.04$ dex, confirming the metallicity is marginally lower than previous studies ($-0.66\pm0.12$, \citealt{alves-brito}, $-0.70\pm0.03$, \citealt{carretta}). 

The stars 20885 and 29861 are also part of the sample of \citet{koch} (star 3 and 1 in their work respectively). We find good agreement between the fundamental parameters found in their study and our results. We find differences in \teff\ of [$-55$\,K,$-36$\,K]; in \logg\ [0.09,-0.11]; in \vmic\ [0.2,-0.04] and in \feh\ of [-0.08,-0.09], which should be compared to our uncertainties of 80K/0.2/0.1/0.06. Thus both \logg\ and \teff\ agrees well within the uncertainties. The higher \vmic\ found for 20885 is likely related to differences in the line selection. We note that in this work as well as the work of \citet{koch}, this star represents an outlier in terms of \vmic. For comparison, \citet{koch} quote uncertainties of 40\,K/0.2\,dex/0.1\kms/0.04\,dex and note that they did not enforce ionization equilibrium, which may serve to explain some of the differences.

We also have three stars in common with the work of \citet{alves-brito} (4794, 5265 and 5968; M8, M11 and M12 in their work). Good agreement is found for \teff\ (mean difference 53K) but substantial differences are found for \logg. Compared to their work we find \logg\ lower by 0.33, 0.9 and 0.6 dex respectively, which will have a significant impact on the derived abundances of elements derived from pressure sensitive lines. As the authors used essentially the same method for deriving the parameters, and did also publish their EW measurements, we did a line-by-line cross-comparison to investigate this issue. This was done for both \feone\ and \fetwo. For all three stars we found that our measured EWs of \feone\ was slightly higher than the comparison work, with median offsets of 0.75, 0.5 and 0.8\,m\AA\ for stars 4794, 5265 and 5968 respectively. For \fetwo\ we found corresponding differences of -0.4, 1.9 and 0.9 m\AA. However, individual differences were found to be as large as 13m\AA\ in a few cases. We are able to reproduce the results of \citet{alves-brito} for the \fetwo\ lines, but this requires placing the continuum level significantly lower than can be justified from our spectra.
 
If the continuum level tends to be underestimated, it will result in a lower EW, which, in turn, is reflected in a higher value of \logg. We use the same \loggf\ values for \fetwo, as the comparison study, but for \feone\ we adopt the values from the GES line list which, on average, are higher than what used by \citet{alves-brito}. We attribute the main source of the discrepancy to the different continuum placements and the different \loggf\ values applied for \feone. Minor differences will also arise from the use of different model atmospheres (ATLAS9 solar-scaled vs. ATLAS12 $\alpha-$enhanced) as mentioned previously.

\begin{table}%
\centering
\caption{Fundamental atmospheric parameters for our stars.}
\begin{tabular}{rcccrr}
\hline
\input{fund_params.tex}
\end{tabular}
\label{fundam}
\end{table}

\subsection{The abundance pattern}

In Fig.~\ref{abupattern} we present the full abundance pattern for our sample stars. The boxes gives the interquartile range, incorporating 50\% of our measurements, with the horizontal line indicating the median value. The whiskers indicate the total range of our measurements, or extends to 1.5$\times$ the second and third quartile range for elements where the full range is larger than this. Any measurements deviating by more than this amount is shown by an open circle. The full list of all measured abundances ratios is also given in Table~\ref{abundancetable}. 

\begin{table*}%
\centering
\caption{Abundances for all elements measured in our stars as well as the total error and number of lines used. The full table is available in the online version of the paper.}
\begin{tabular}{rccccccccccccccc}
\hline
\input{abundance-ratios-cut-v7.tex}
\end{tabular}
\label{abundancetable}
\end{table*}

\subsubsection{Light elements}
\label{polluters}
The proton-capture elements Na and Al can be created in the hydrostatic burning in the cores of massive stars during the main sequence phase, as well as at the base of the convective envelope in AGB stars, via the so called Hot Bottom Burning mechanism \citep{renzini}. Thus, these elements act as important tracers of the burning conditions required by the polluters, in order to create the abundance patterns observed today.  

In essentially all GCs where these elements have been studied, large spreads of Na have been observed (\eg\ \citealt{carretta,gratton} and \citealt{carretta3}). Spreads of Al have also been reported for a number of clusters, albeit this is not as common as the Na variation. With respect to sodium, 47 Tuc is no different, and we find a spread in [Na/Fe] of $\sim$0.5\,dex, in the range $[0.01;0.50]$ dex, which falls within the range reported by other studies \citep{cordero,alves-brito}. That the spread is significant compared to the uncertainties on the abundance ratios themselves can be seen by inspecting Table~\ref{spread} where we give the interquartile range (IQR) for all elements, as well as the median uncertainty on the measurements. We choose the IQR as a measurement of the spread, since this is more robust towards outliers and potentially skewed data than the standard deviation. We also give the mean value of the abundance ratios.

\begin{table}%
\centering
\caption{IQR, mean value and median uncertainty for all measured elements.}
\begin{tabular}{lccc}
\hline
\input{abundancedispersion_IQR_v2.tex}
\end{tabular}
\label{spread}
\end{table}

\begin{figure*}%
\centering
\includegraphics[width=\textwidth,trim= 2cm 13cm 2cm 1cm, clip=true]{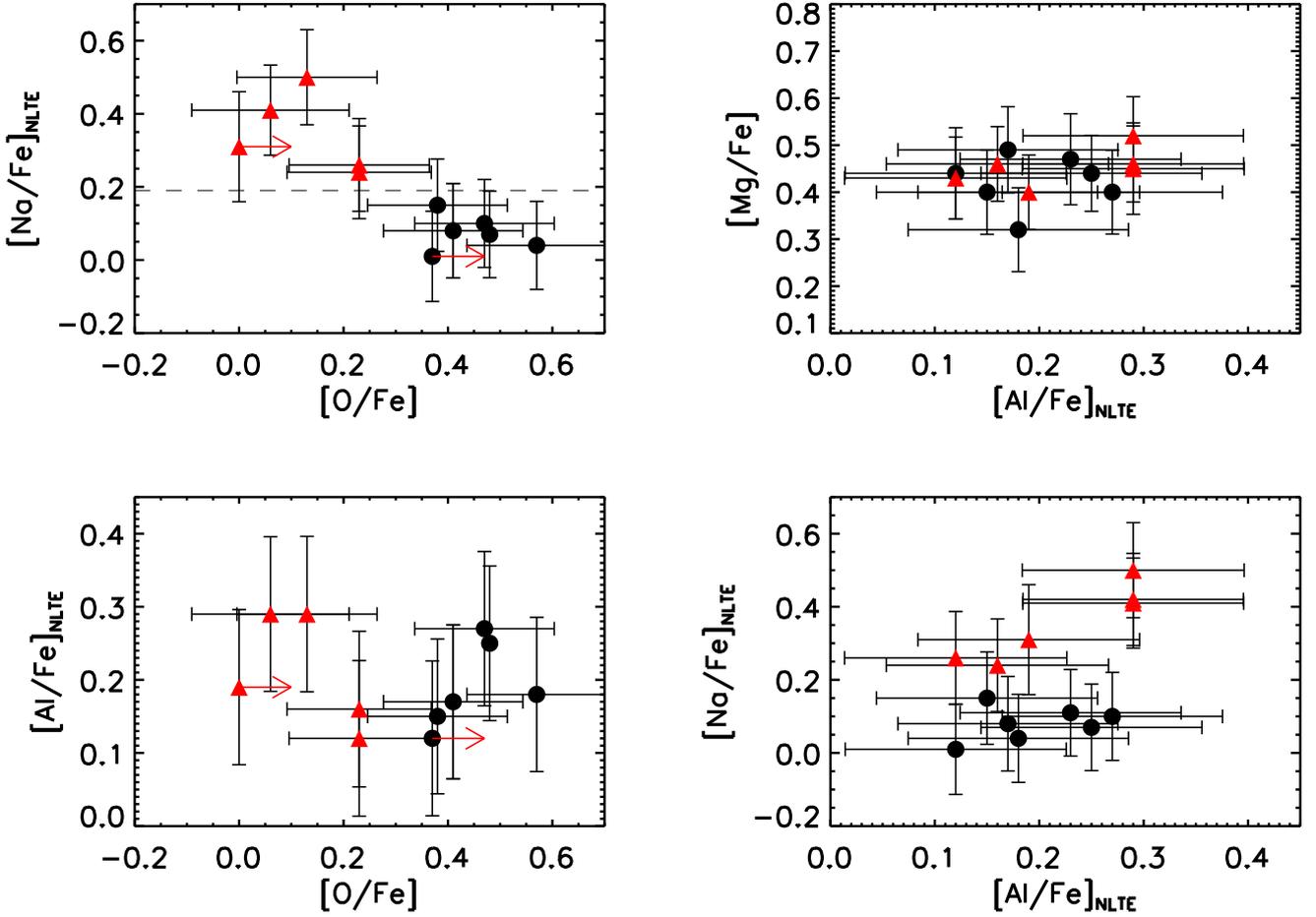}%
\caption{Correlations between the light elements. Top left: [Na/Fe] vs. [O/Fe], top right: [Mg/Fe] vs. [Al/Fe], bottom left: [Al/Fe] vs. [O/Fe], bottom right: [Na/Fe] vs. [Al/Fe]. The dashed line indicates the adopted cut between the pristine (black circles) and polluted (red triangles) population of stars.}%
\label{lightelem-ratio}%
\end{figure*}

The abundance of Na is usually observed to be anti-correlated with that of oxygen. Inspecting Fig.~\ref{lightelem-ratio}, we confirm this behaviour for our stars in 47 Tuc. The two stars with red arrows are stars 5265 and 6798, where telluric emission is present in one wing of the oxygen lines, so the derived [O/Fe] should be considered as a lower limit only. In our case, the anti-correlation is much more pronounced than seen in the study of \citet{koch}, who only observed a small scatter with no clear correlation. However, a clear anti-correlation was recently reported by \citet{cordero}, who analyzed a sample of more than 160 giants in 47 Tuc. Our range in both elements is fully consistent with this study. 

\begin{figure}%
\centering
\includegraphics[width=\columnwidth, trim= 1.5cm 0cm 1.5cm 0cm, clip=true]{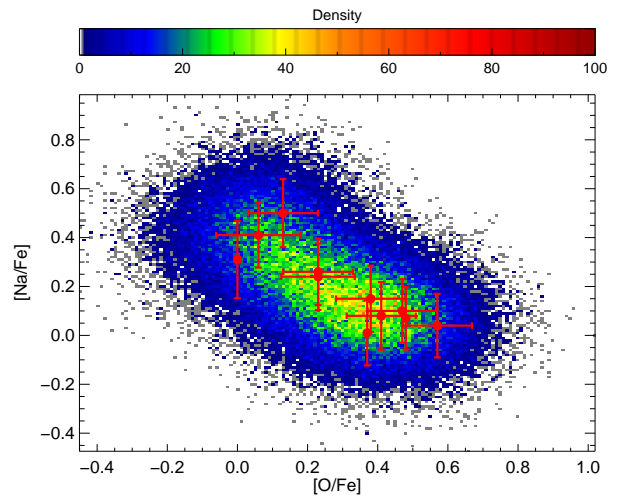}%
\caption{The distribution of [Na/Fe] vs. [O/Fe] from our bootstrapping exercise, shown as a density plot. Bin-size is 0.01. The measurements are shown in red. The two targets with no oxygen errorbars only got lower limits, and were excluded from the bootstrapping calculations.}%
\label{density}%
\end{figure}

Computing the Kendall $\tau$ correlation coefficient for the set of measurements yields $\tau=-0.88$, indicating a strong anti-correlation with a statistical significance of 0.001, allowing us to reject the null hypothesis of no correlation with more than 99\% confidence. In order to quantify the strength of the correlation, we performed a bootstrapping exercise, changing the value of both abundance ratios by adding perturbations to the measured values. The perturbations were drawn from a Gaussian distribution with mean zero and standard deviation corresponding to the total uncertainties of our measurements. The two stars with only lower limits of the oxygen abundance were excluded from this exercise. 10,000 realizations were made for each abundance ratio, and for each set of ([Na/Fe], [O/Fe]) the Kendall $\tau$ was computed. The distribution of the abundances is shown in Fig~\ref{density}, using a bin-size of 0.01. For convenience, we overplot the actual measurements, including the stars with only lower limits. Inspecting the distribution of the $\tau$ values, they were found to follow a normal distribution. Fitting a Gaussian, we computed the mean value of the Kendall $\tau$, finding $\tau=-0.60\pm0.15$, so a weaker correlation is found from this exercise.

The pristine population of stars in 47 Tuc should be indistinguishable from field stars at the same metallicity and it is thus possible to separate the polluted stars from the pristine, using their [Na/Fe] ratio. We adopt a comparison sample from the recent study of abundances in 714 stars in the solar neighbourhood by \citet{bensby}, although we note that their their sample
consists primarily of dwarf stars whereas our program stars are all giants. The Na abundances from the \citet{bensby} study has been corrected for NLTE effects using the same source as we applied for our stars. We take the mean of their [Na/Fe] measurements for stars with $-0.9<\rm {[Fe/H]}<-0.6$ as a measure of the typical value of [Na/Fe] for stars in the field with a metallicity comparable to 47 Tuc. We find [Na/Fe]$_{{\rm field}}=0.09\pm0.05$ dex. Taking the mean value plus twice the standard deviation as an upper limit of the [Na/Fe] ratio in typical field stars, we consider stars with $\rm {[Na/Fe]}\geq0.19$ dex to belong to the polluted population of stars. This approach allow us to separate the stellar populations in 47 Tuc, even if we are not sampling the full range of the Na variations. The stars identified as belonging to the polluted population are indicated by red triangles in Fig.~\ref{lightelem-ratio} as well as in all other relevant figures. We also identify them by boldface numbers in all relevant tables.

Turning our attention to the measured aluminium abundances in Fig.~\ref{lightelem-ratio}, it is evident that they show very little spread, with a mean and standard deviation of [Al/Fe]$_{mean}=0.21\pm0.06$\,dex. This should be compared to the uncertainty of the individual [Al/Fe] determinations, which are on the order of $\pm0.11$\,dex, so our results are consistent with a single value for our sample of stars. This is in contrast to the the results presented by both \citet{carretta2009} and \citet{cordero}, who reported a significant variation in their measured Al abundances.

The fact that we do not observe any spread in Al in our sample of stars is a consequence of treating the Al synthesis in NLTE, rather than relying on a standard LTE analysis. If the Al lines are treated in LTE, we observe a larger range of abundances, as will be discussed in more detail in Sect.~\ref{al-nlte}. The disappearance of the Al variation in this case, is more related to the small number statistics of stars in our sample, combined with the spread in our stellar parameters. It does not imply that no Al variation exists within the cluster. The presence of a variation is evident from previous studies by \eg\ \citet{carretta2009} and \citet{cordero}, who report significant variations in Al for stars with identical parameters. Because NLTE corrections are governed by the stellar parameters, such variation would not disappear as a consequence of a full NLTE analysis of their stars. In our case, the sample simply does not cover the full range of Al variations for a given set of stellar parameters.

The two bottom panels in Fig~\ref{lightelem-ratio} show [Al/Fe] vs. [O/Fe] and [Na/Fe] vs. [Al/Fe]. No correlations are visible in either of the two plots, in accordance with our finding that [Al/Fe] for our stars are consistent with a single value. This also indicates that, at least for the stars in our sample, the polluted generation has been enriched by material from stars where the Mg-Al burning cycle was not activated to any significant extent.

\subsubsection{$\alpha$-elements}
The $\alpha$-elements (O, Mg, Si, Ca and Ti are enhanced in ATLAS12 models) show a roughly constant behaviour, with a mean enhancement of $0.34\pm0.04$\,dex relative to iron, in good agreement with what is found by other studies of 47 Tuc (\eg\ [$\alpha$/Fe$]= 0.41$\,dex; \citet{koch}; 0.29\,dex, \citet{cordero}; 0.3\,dex, \citealt{alves-brito}). Inspecting Table~\ref{spread}, oxygen is the only exception, as it shows a variation and the well-known anti-correlation with Na as discussed earlier. Excluding oxygen from the calculation of the $\alpha$-enhancement, we find a mean value of [$\alpha$/Fe$]=0.35\pm0.03$\,dex, so an insignificant change compared to using the entire set of abundances.
We note that Ti exhibits a large scatter as well as discrepancies between the neutral and ionized species, which is a well-known issue that still has not been entirely resolved, although improvements were clearly made with the recent update of the log(gf) values by \citet{lawler}. The remaining part of the discrepancy can likely be attributed to NLTE effects on Ti\,I as investigated by \citet{bergemann2} who showed that the abundance from lines of the neutral species are systematically underestimated in LTE. Following the recommendations in her work, we adopt the Ti\,II abundances as the best measure of Ti in the stars, as the ionized species have negligible NLTE corrections. Further, in a high-metallicity environment like 47 Tuc, the useable Ti lines are very strong and enters the saturated part of the curve-of-growth, making them very weakly sensitive to changes in abundances. 

\begin{figure*}%
\centering	
\includegraphics[width=\textwidth,trim= 2cm 4cm 0cm 2cm, clip=true]{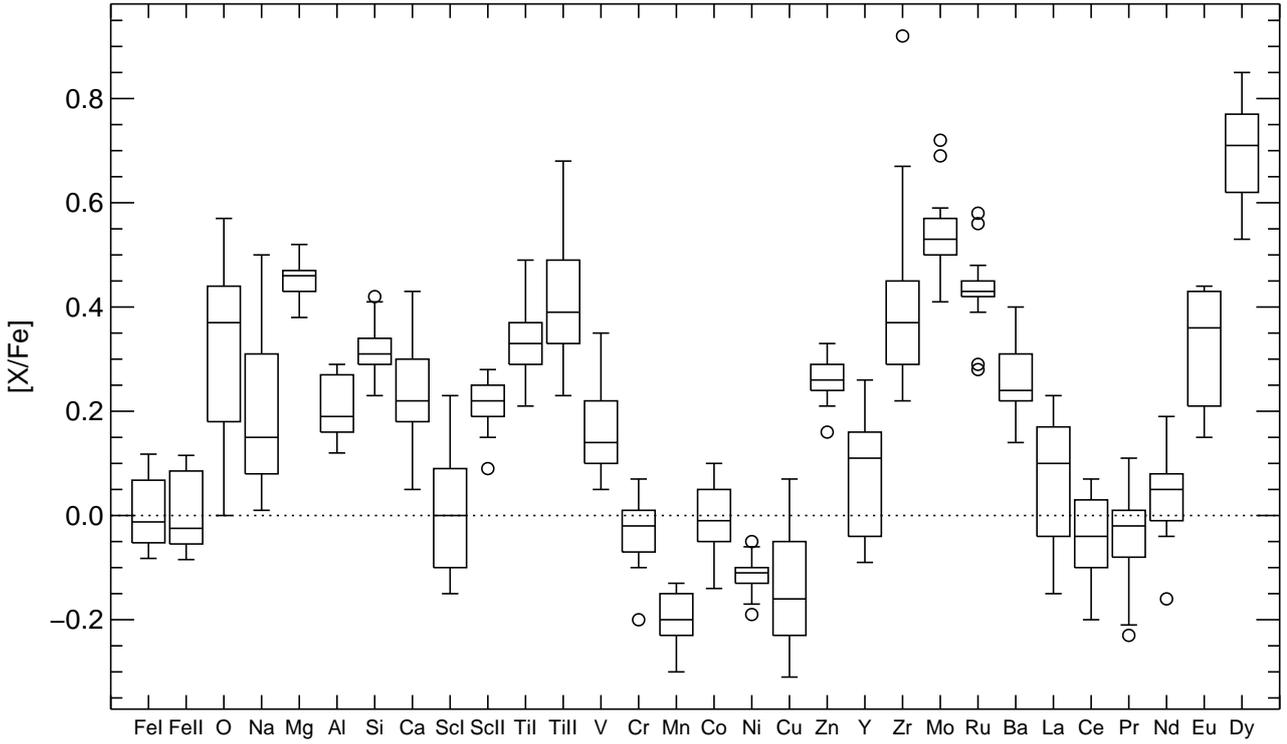}%
\caption{The abundance pattern for our sample stars. Shown are the interquartile ranges. The most extreme outliers are indicated with open circles.}%
\label{abupattern}%
\end{figure*}

Scandium is intermediate between pure $\alpha$ elements and iron-peak elements and the formation processes is not entirely clear at present. Inspecting Fig.~\ref{abupattern}, a clear offset is seen between the Sc\,I and Sc\,II abundances, with the median values differing by about 0.2\,dex. This is most likely related to NLTE effects influencing the lines of the neutral species. Whereas this effect has not been studied in giants, \citet{zhang} found for the Sun that NLTE corrections are positive and on the 0.2\,dex level for the neutral species, bringing Sc\,I and Sc\,II back into agreement. Whereas not directly applicable to our sample of stars, their study at least suggests that NLTE effects could be responsible for the observed disagreement.

\subsubsection{Iron-group elements}
Each of the iron group elements (V, Cr, Mn, Fe, Co, Ni) are found to be constant and we see no evidence for an intrinsic spread for any of them (see Table~\ref{spread}). Mn and Ni appears to be slightly underabundant compared to Fe, whereas Co is found to be overabundant. This is seen also in studies of other GCs like \eg\ M75, \citep{kacharov}; M71, \citep{boesgaard}; NGC 1851, \citep{carretta2}; M4 and M5, \citep{yong} as well as in globular clusters in the Large Magellanic Clouds \citep{colucci}.

\subsubsection{Neutron-capture elements}
All elements beyond Ni are produced by neutron captures, either through the rapid(r)-process associated with a very high neutron flux during neutron star mergers or SN II explosions, or, alternatively, through the slow(s)-process, taking place on longer timescales in e.g. AGB stars. Many elements can be created through both processes, but a few can be associated largely with one of the two mechanisms. The s-process elements Y and Zr both show a scatter that does not appear to be correlated with neither stellar parameters nor the light element abundance variations (See Fig~\ref{NaVZr} in the appendix). However, the IQRs for [Zr/Fe] and [Y/Fe] are 0.16 dex and 0.20 dex, which should be compared to typical uncertainties of 0.19 and 0.18 dex respectively, for individual stars. The abundance dispersion is thus consistent with a delta function convolved with the measurement uncertainty, i.e., there is no evidence for an abundance variation given the measurement uncertainty (also see Table~\ref{spread}). Lines of both elements have been treated in LTE. We note that while NLTE effects on Zr-I lines may influence our derived abundances, given the narrow range in stellar parameters of the program stars, it is unlikely that neglect of NLTE effects masks the presence of a genuine abundance spread for Zr 	in 47 Tuc. A recent study by \citet{velichko} indicates that both ionization stages of Zr are influenced by NLTE, in particular Zr\,I. They found corrections as large as 0.3 dex, increasing with decreasing \teff, \logg\ and metallicity. Their study only treated the 4241\,\AA\ and 4687\,\AA\ Zr\,I lines and since NLTE corrections can be strongly line-dependent, the potential NLTE corrections of the lines used in this work may differ from theirs by a non-negligible amount. 

The heavy s-process elements La and Ba show no correlation with any of the light elements that exhibits the abundance variations, being fully consistent with just a single value (See Fig~\ref{NaMoDy} and Table~\ref{spread}). This result suggests that low-mass AGB stars have not contributed strongly to the enrichment of the gas incorporated in the polluted population of stars, as these elements are predominantly produced in lower mass AGB stars \citep{straniero}.

The elements Mo, Ru, Pr and Nd have almost equal contributions from both slow and rapid neutron capture processes \citep{bisterzo}, so it is difficult to interpret their abundances in terms of polluter candidates. None of these elements exhibit any intrinsic scatter nor correlations with the light elements. We note that Mo and Ru appear enhanced, compared to Pr and Nd, which is seen in some GCs where these elements have been studied (M75, \citealt{kacharov}), whereas \citet{roederer} found the opposite behaviour in their study of M22. This cluster also shows a spread in iron, so the reason for this behaviour may as well be linked to the fact that the different populations have seen additional SNe contributions to their abundance patterns, which is not expected to be the case for 47 Tuc. Using Mo and Ru to discriminate between different polluters is further complicated by the complex formation channels for this element (see \citealt{hansen}).

The r-process element Eu is found to be constant within the measurement uncertainties (see Table~\ref{spread}) and whatever small variation is observed is also not correlated with the light element variation (Fig~\ref{NaMoDy}). Since Eu is almost exclusively produced in the r-process (94\%, \citealt{bisterzo}), this element is often used as a measure of the overall r-process enrichment. When comparing Eu to the s-process dominated species Ba (85\% s-process, \citealt{bisterzo}), we find a value of $\log\epsilon({\rm Ba/Eu})=1.59\pm0.17$, which is close to the solar value of 1.66 based on the \citet{arlandini} values, suggesting that the neutron capture elements in 47 Tuc have seen a prevalent s-process contribution, as the pure r-process value is 0.97 for the same abundance ratios. This is at odds with what was found by the study of \citet{cordero}, who compared Eu to La (76\% s-process, \citealt{bisterzo}), suggesting that the neutron capture elements in 47 Tuc were r-process dominated. Indeed, if we use the same element ratios as in their study, we find $\log\epsilon({\rm La/Eu})=0.33\pm0.04$, which would suggest a larger r-process contribution relative to the s-process, as the pure r-process ratio is 0.27, compared to the solar ratio of 0.69 \citep{arlandini}. However, both La and Ba indicate that 47 Tuc has seen some s-process contribution, although the amount differs. This is also in agreement with what was seen by \eg\ \citet{alves-brito}, who found [Ba/Fe]\,$=0.31\pm0.07$\,dex to be significantly higher than their [La/Fe] ratio of $0.05\pm10$.

We decided not to use Dy as an s-process probe (84\% s-process, \citealt{bisterzo}), as this element appears enhanced compared to any other element, and in particular it is enhanced with respect to Eu. Dy has only been measured in clusters in a few cases. A mild enhancement relative to Eu was found in both M75 \citep{kacharov} and M22 \citep{roederer}, whereas \citet{carretta2} found [Dy/Fe] to be equal to [Eu/Fe] in the GC NGC 1851. \citet{cohen}, on the other hand, found that Dy was depleted relative to Eu in both M3 and M13, so the behaviour of this element is unclear at present, but may be cluster dependent. Finally, we note that \citet{roederer2} argues, albeit in a different metallicity regime, that measuring abundance ratios of Pb is the only definitive way to examine whether a s-process contribution has occurred within a stellar population.

\section{Discussion}
The majority of our results show good agreement with previous studies of 47 Tuc, but a few points warrant a deeper discussion.

Several studies (\citealt{brown}, BW92, \citealt{james2004}, J04, \citealt{alves-brito}, AB05, \citealt{wylie}, W06, \citealt{mcwilliam}, McW08) have reported abundances of the s-process peak elements, Y, Zr and Ba, but the disagreement between the individual studies is significant. This is clearly seen in Table~\ref{yzr}, where we quote the mean values from the relevant studies as well as the values found in this work, given as the mean $\pm$ the standard deviation. The reason for the disagreement of the [Y/Fe] ratios can easily be understood, since the Y\,II lines used to derive the yttrium abundances suffer from HFS. If this is not properly taken into account, the abundance will be overestimated. The studies of \citet{brown} and \citet{wylie} are not applying a HFS analysis, and thus they find a higher abundance than any of the other quoted studies. Our mean abundance is slightly higher than the two remaining studies which reported Y abundances. A direct comparison to the \citet{james2004} study is difficult, as they use a different set of lines which fall outside our spectral range. They also found an offset in the abundance between main sequence turn-off and subgiant stars, with the former having an [Y/Fe] of $+0.06\pm0.01$, in better agreement with our measurements. No explanation for this offset was given, but we chose to compare our results to the measurements of subgiants, as these are more similar to our sample than the dwarf stars yielding the higher abundance ratio. \citet{mcwilliam}, on the other hand, use integrated light spectroscopy and derive their abundance based on only a single line, making a comparison to our results difficult, although one would expect the mean derived from multiple stars to be comparable to what would be measured from integrated light, assuming that the integrated light lines can be synthesized correctly.

\begin{table}%
\centering
\caption{[Y/Fe], [Zr/Fe] and [Ba/Fe] as reported by several other studies as well as what is found in this work.}
\begin{tabular}{lccc}
\hline
Source & [Y/Fe] & [Zr/Fe] & [Ba/Fe] \\
\hline\hline
BW92 & $+0.48\pm0.11$ & $-0.22\pm0.05$ & $-0.22\pm0.12$ \\
J04 & $-0.11\pm0.10$ & $-$  & $+0.35\pm0.12$ \\
AB05 & $-$ & $-0.17\pm0.12$ & $+0.31\pm0.07$ \\
W06 & $+0.65\pm0.18$ & $+0.69\pm0.15$ & --  \\
McW08 & $-0.13$ & $+0.05$ & $+0.02$ \\
\hline
This work & $+0.09\pm0.11$ & $+0.39\pm0.20$ & $+0.28\pm0.07$\\
\hline
\end{tabular}
\label{yzr}
\end{table}
	
Turning our attention to the [Zr/Fe] ratio, again there is no clear picture of the actual abundance ratio. The study of \citet{alves-brito} relies on the synthesis of the 6143\,\AA\ line and they do not report taking HFS into account in their analysis. Thus it is surprising that we find a higher abundance, even when performing a HFS analysis. We do note, however, that the 6143\,\AA\ line yields a systematically lower abundance, compared to the two additional lines we used, but never more than 0.1 dex, which is insufficient to reconcile our measurements with those of \citet{alves-brito}. Inspecting our abundances on a star-by-star basis we never see line-to-line scatter larger than 0.1 dex, suggesting that our abundance measurements are internally robust. However, as can be seen from Table~\ref{abuuncert}, the Zr abundance is very sensitive to changes in the \teff\ scale, which can explain part of the difference found. Comparing directly the three stars in common between our sample and the study of Alves-Brito, the most deviant star in terms of parameters (5265) would have its Zr abundance lowered by $\sim$0.2 dex if the parameters are changed to force agreement with the Alves-Brito study. This can explain part of the disagreement. The remaining disagreement can likely be explained by a combination of differences in continuum placements, as discussed earlier for the iron lines, and the use of different model atmospheres (see Table~\ref{abumodels}). Our results are in somewhat better agreement with the MW08 study and we note that we use the same atomic line data as in their study, although we still find a significantly larger Zr abundance. 

Our results for Barium compare well with what has been found by most other studies, although we find a higher abundance than the McWilliam study, which again might be related to their use of integrated light spectroscopy. Also, we are using slightly different atomic parameters, which will also lead to differences in the derived abundances. 

\subsection{NLTE effects for Al}
\label{al-nlte}
As mentioned previously, we found that NLTE was relevant for Al and sensitive to the stellar parameters, which may introduce problems with abundances derived from a LTE analysis. Since Al is a key element for constraining possible polluter candidates in GCs, this is important to take into account. Significant NLTE effects on Al lines, even at high metallicity have been reported by \citet{gehren,gehren2}, who studied the effect in a large sample of dwarf stars, showing that the effects were non-negligible even at solar metallicity. We confirm this behaviour in our giants, although the corrections are in the opposite direction of that in the dwarf stars.

\begin{figure*}%
\centering
\includegraphics[width=\textwidth,trim= 2cm 13cm 0cm 6cm, clip=true]{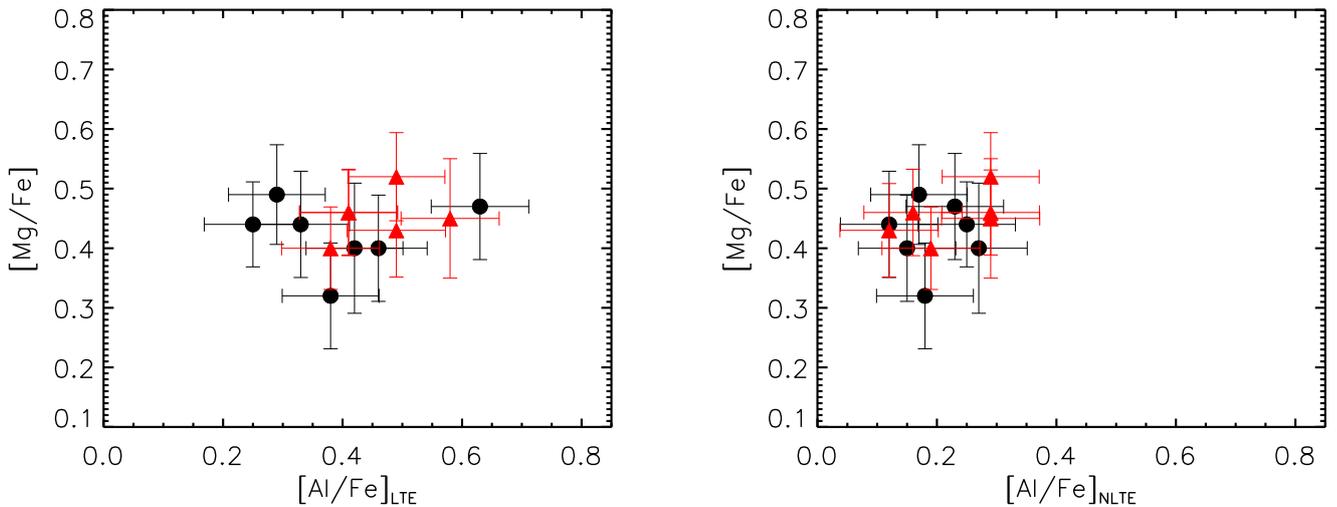}%
\caption{[Mg/Fe] vs. [Al/Fe] in LTE (left) and NLTE (right). The observed range in [Al/Fe] decreases when NLTE is taken into account.}%
\label{mg-al-corr}%
\end{figure*}

A variation in the Mg and Al abundances has been reported in more than 20 GCs, with a large part (18) of the data coming from the study of bright giants by \citet{carretta2009}. However, the Al abundances in the literature are all derived under the assumption of LTE and often based on a small number of bright stars. Inspecting Fig.~\ref{mg-al-corr} where we plot both the LTE and NLTE values of [Al/Fe] it is immediately apparent that the total range of the [Al/Fe] ratio for our sample decreases by more than a factor of two when treated in NLTE, making our results consistent with only a single abundance across our sample of stars. As discussed previously, this is related to our sampling of stars in the CMD, underlining the importance of dense sampling of the parameter space in order to unambiguously claim a variation in Al. Unless one samples stars with near-identical parameters one may overestimate the actual spread in [Al/Fe] in LTE, in particular if the extremes in [Al/Fe] coincide with extremes in stellar parameters. In the case of 47 Tuc, however, significant variations in Al has been observed in stars with identical parameters \citep{carretta3,cordero} so this cluster exhibits a genuine spread. The latter study use the 6696\,\AA\ and 6698\,\AA\ lines, so it is directly comparable to ours, whereas \citet{carretta3} uses the NIR lines in the region around 8773\,\AA, which may have different NLTE corrections. But even allowing for corrections of a different magnitude for the NIR lines would not result in a disappearance of the observed variation in their study. Another effect of treating the aluminium lines in NLTE is an overall shift to lower values as can also be seen by comparing the LTE and NLTE results in Fig.~\ref{mg-al-corr}. This result will be valid in general and implies that the overall enhancement of Al relative to iron may be overestimated in earlier works. However, this needs to be investigated in a larger sample of stars and ideally in multiple GCs to determine the true extent of this.

Interestingly, the stars in our sample do not separate clearly in Fig.~\ref{mg-al-corr}. One would expect that stars which are enhanced in Al and depleted in Mg, would also be enhanced in Na and depleted in O, as the Mg-Al anti-correlation is created by a more advanced burning stage. However, a variation in Na-O does not imply a variation also in Mg-Al. We see no evidence for the Al abundance to be correlated with neither O nor Na, in our sample of stars (See Figs.~\ref{mg-al-corr} and ~\ref{NaOTi}). Depending on the polluter candidate, it is certainly possible to create an Na-O anti-correlation, without activating the Mg-Al burning cycle, responsible for the anti-correlation of the latter (See \eg\ \citealt{decressin,bastian}). This is also a well-known observational fact, where only a subset of the globular clusters show significant variations in Al (see \eg\ \citealt{carretta2009}). Indeed, \citet{cordero} also do not find a correlation between Na and Al, in their much larger sample of stars in 47 Tuc (See their Fig. 8), but merely finds a variation in Al for any given Na abundance, consistent with our findings here.

\subsection{Comparison with M71}
The GC M71 has a [Fe/H]\,$=-0.71$ \citep{ramirezM71}, and is thus a close match to 47 Tuc in terms of metallicity but not nearly as massive. Also for this cluster, only a small dispersion is seen in the [Mg/Fe] ratio ($\Delta{\rm [Mg/Fe]}=0.11$), very similar to what we observe for our sample of stars (\citealt{ramirezM712} and \citealt{melendez2}), suggesting that the smaller variation observed may be related to the high metallicity of the cluster. Our values of [Mg/Fe] are offset by around 0.2\,dex, compared to \citet{melendez2}, whereas better agreement is seen with the range reported by \citet{ramirezM712}. Both studies also found only a small range in aluminium abundances ($\Delta{\rm [Al/Fe]}=0.16$), which is comparable with what we observe, but we note that 47 Tuc certainly exhibits variations larger than what we found, as discussed above. 

Inspecting the behaviour of the elements heavier than iron, we again see a good agreement with the observed abundance pattern of M71 \citep{ramirezM712}, with the exception of Y and Zr, where we find an overabundance relative to iron, opposite what is reported by \citet{ramirezM712}. However, the more recent study of \citet{melendez2}, do report an overabundance of Zr, albeit lower than what we find for 47 Tuc. This may indicate that 47 Tuc has seen a stronger contribution from the weak s-process than M71. Regarding La and Ba, M71 appears to be slightly more enhanced in these elements than 47 Tuc, in particular La, indicating that even though the two clusters have essentially the same metallicity, they have not undergone identical evolutions, which is not surprising considering the mass difference between the two clusters. We note that the same trend of Ba being enhanced to the level of Eu as well as being more enhanced than La is seen in both cases.

\section{Conclusions}
Using observations of unprecedented quality, we have derived abundances for a total of 27 elements in the globular cluster 47 Tucanae, spanning the range from O to Dy. We confirm the known anti-correlation of Na and O, whereas we find no indication of an  anti-correlation between Mg and Al, which has previously reported by \eg\ \citet{carretta2009}, but this is a consequence of not sampling the full extent of the [Al/Fe] enhancement in our sample of stars. The variation of Na and O confirms that at least two populations of stars are present in 47 Tucanae, in line with what is seen in studies of both photometry as well as spectroscopy. We find a mean $\alpha$-enhancement of [$\alpha$/Fe]\,=\,$0.34\pm0.03$. For the overall metallicity we find a value of \feh\,$=-0.78\pm0.07$ which is in good agreement with other recent studies of 47 Tuc. Inspecting the iron-peak elements, we see no indication of a variation, within the measurement uncertainties, consistent with a mono-metallicity. The same holds for the s-process and r-process dominated species. Inspecting abundance ratios of the s-process dominatied species (Ba, La) to that of Eu, we see indications that 47 Tuc has seen some s-process contribution to the abundances. Finally, abundance of the trans-iron elements Mo, Ru, Pr and Nd are found to be constant across all stars in our sample to a very high degree. We also do not observe any statistically significant correlations between [Na/Fe] and any other element, besides [O/Fe], consistent with the interpretation that no intrinsic variation is present in our sample of stars.

The use of a full NLTE synthesis of the Ba and Al lines eliminated spurious trends with \teff\ found for both species, and it reduces the overall scatter significantly. We found that the observed range in [Al/Fe] decreased by a factor of $\sim$two, with a value consistent with a single abundance across our sample of stars. We note that this is due to our small sample of stars, which is not covering the full range of Al abundances in the cluster and should not be taken as an indication that no Al spread exits in 47 Tuc. Further, the overall enhancement of Al was also found to decrease compared to the abundances derived from an LTE analysis. If confirmed in other clusters, this may impact the constraints put on the polluter candidates. Since we only observe a small sample of giants here and this issue will need to be studied in a larger number of stars and in different clusters before any firm claims can be made regarding this.

These results will help constrain the mechanisms for creating the light element abundance variations. A detailed interpretation of the results in terms of consequences for the intra-cluster polluter candidates will be presented in Paper II, which will also add an analysis of the isotopic mixture of magnesium to the already derived abundances.

\begin{acknowledgements}
AOT, LS and HGL acknowledges support from Sonderforschungsbereich SFB 881 "The Milky Way System" (subprojects A4 and A5) of the German Research Foundation (DFG). LS acknowledges funding from Project IC120009 "Millennium Institute of Astrophysics (MAS)" of
Iniciativa Cient\'{i}fica Milenio del Ministerio de Econom\'{i}a, Fomento y Turismo de Chile. SMA and SAK acknowledges SCOPES grant No. IZ73Z0-152485 for financial support. The work of MA has been supported by the Australian Research Council (grant FL110100012). J.M. thanks support from FAPESP (2010 / 50930-6). This research took advantage of the SIMBAD and VIZIER databases at the CDS, Strasbourg (France), and NASA's Astrophysics Data System Bibliographic Services. 
\end{acknowledgements}

\bibliographystyle{aa}
\bibliography{47Tuc_paper1_v11} 

\appendix
\section{Light element correlations}
We here present plots of [X/Fe] vs. [Na/Fe] to trace elements that vary in step with the light element variations.
\begin{figure*}%
\centering
\includegraphics[width=\textwidth,trim= 1cm 3cm 1cm 3cm, clip=true]{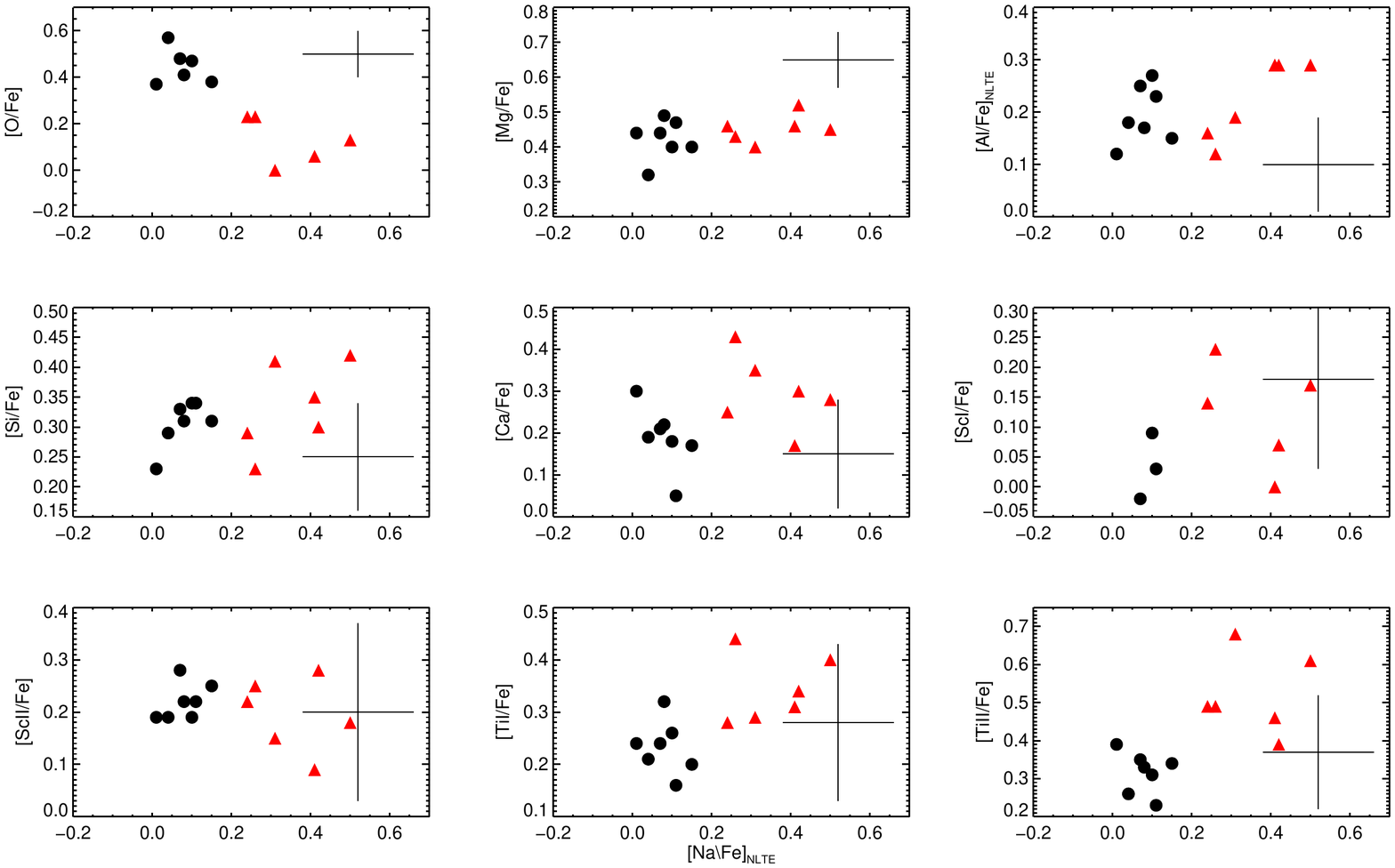}%
\caption{[X/Fe] vs. [Na/Fe] from O to Ti. In each plot is shown representative uncertainties. Symbols the same as in Fig.~\ref{lightelem-ratio}}%
\label{NaOTi}%
\end{figure*}

\begin{figure*}%
\centering
\includegraphics[width=\textwidth,trim= 1cm 3cm 1cm 3cm, clip=true]{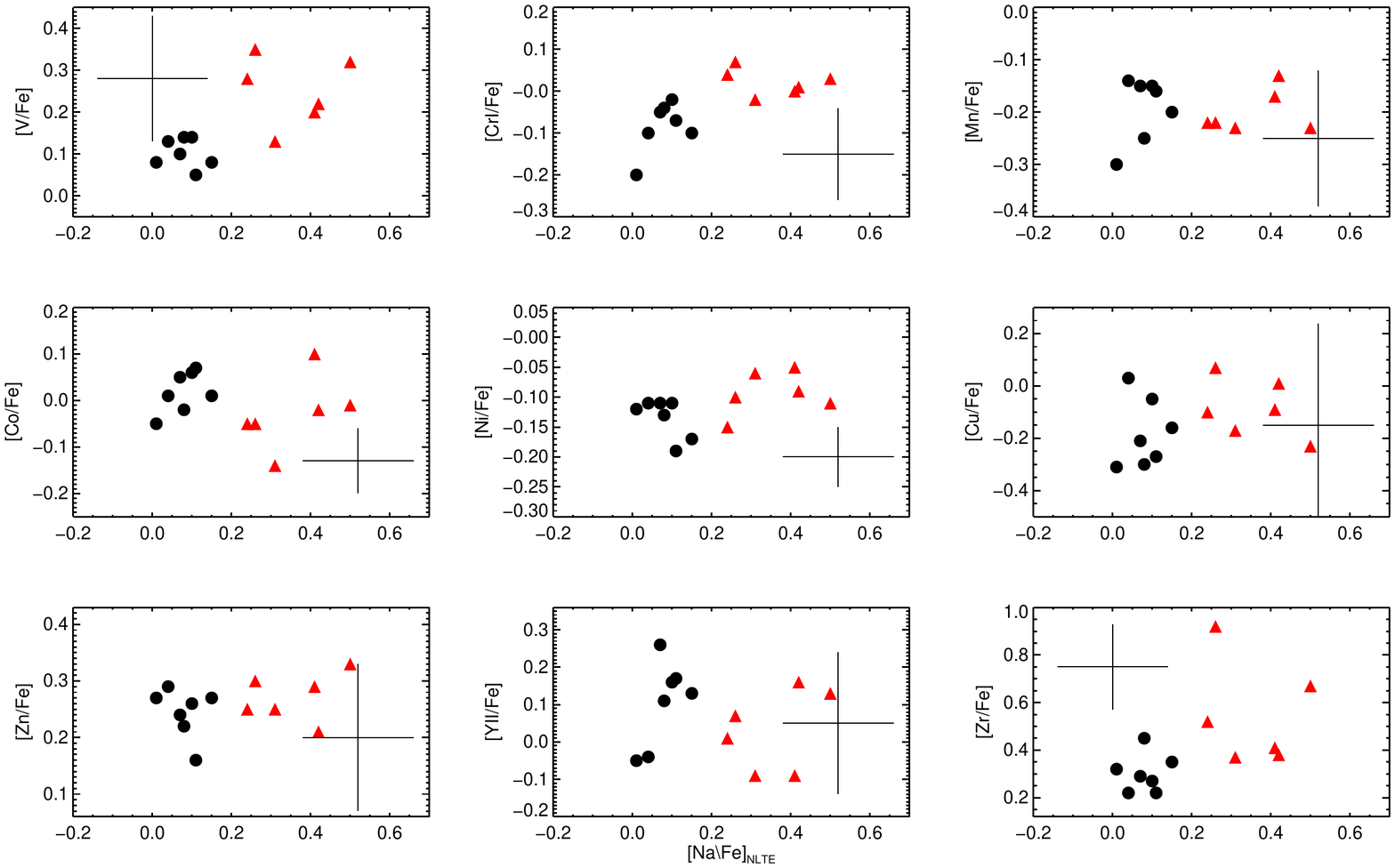}%
\caption{[X/Fe] vs. [Na/Fe] from V to Zr. In each plot is shown representative uncertainties. Symbols the same as in Fig.~\ref{lightelem-ratio}}%
\label{NaVZr}%
\end{figure*}

\begin{figure*}%
\centering
\includegraphics[width=\textwidth,trim= 1cm 3cm 1cm 3cm, clip=true]{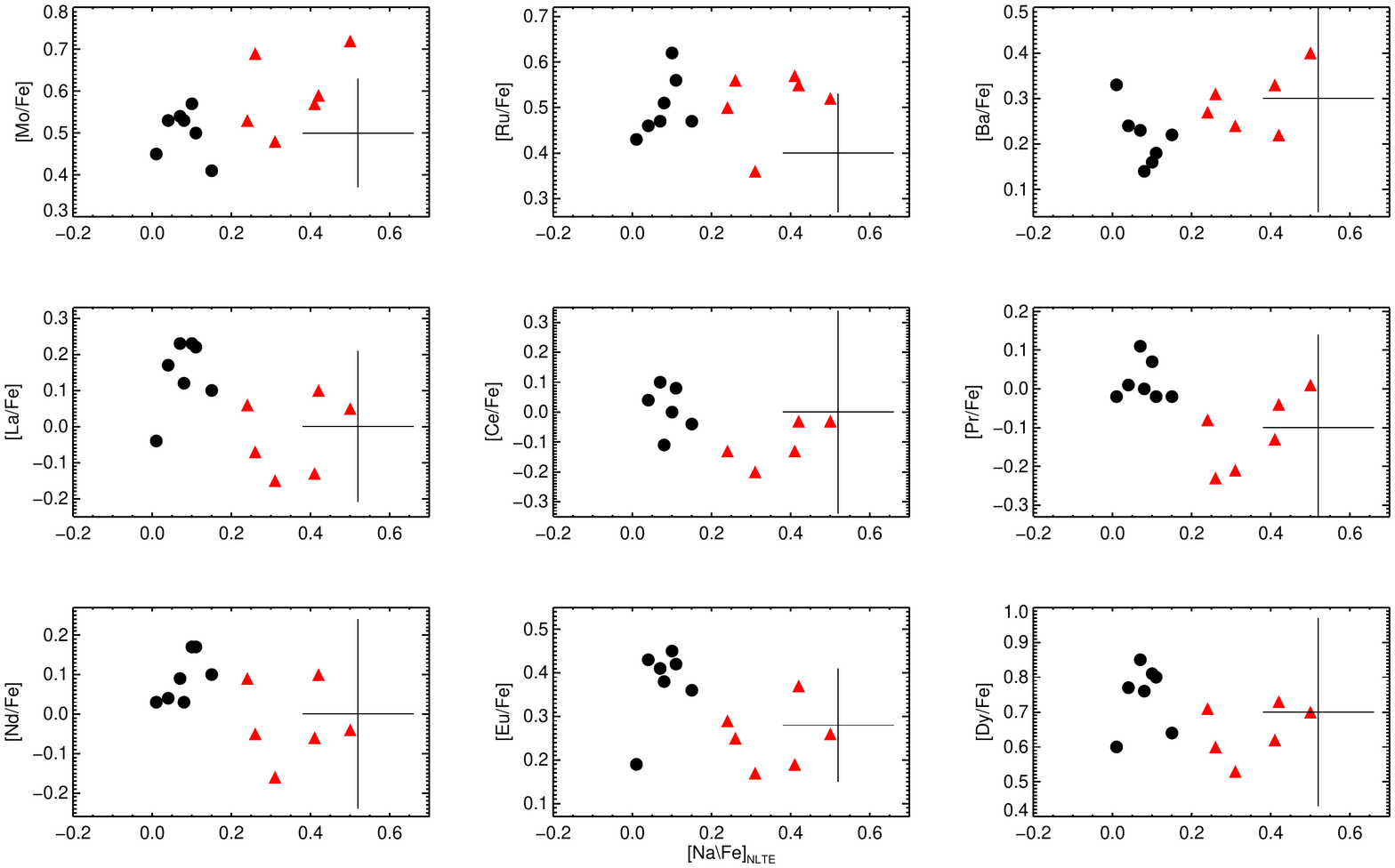}%
\caption{[X/Fe] vs. [Na/Fe] from Mo to Dy. In each plot is shown representative uncertainties. Symbols the same as in Fig.~\ref{lightelem-ratio}}%
\label{NaMoDy}%
\end{figure*}

\end{document}

%% file: non-hfs-lines-cut.tex
6300.304 &  8.0 &  -9.715 & 0.000 \\
6363.776 &  8.0 & -10.190 & 0.020 \\
6154.226 & 11.0 &  -1.547 & 2.102 \\
6160.747 & 11.0 &  -1.246 & 2.104 \\
6318.717 & 12.0 &  -1.950 & 5.108 \\
6319.237 & 12.0 &  -2.165 & 5.108 \\
6319.495 & 12.0 &  -2.803 & 5.108 \\
5557.063 & 13.0 &  -2.11  & 3.143 \\
6696.023 & 13.0 &  -1.48  & 3.143 \\
6698.673 & 13.0 &  -1.78  & 3.143 \\
5690.425 & 14.0 &  -1.773 & 4.930 \\
5701.104 & 14.0 &  -1.953 & 4.930 \\
5948.541 & 14.0 &  -1.130 & 5.082 \\
6131.573 & 14.0 &  -1.556 & 5.616 \\
6131.852 & 14.0 &  -1.615 & 5.616 \\
6142.483 & 14.0 &  -1.295 & 5.619 \\
6155.134 & 14.0 &  -0.754 & 5.619 \\
6155.693 & 14.0 &  -2.252 & 5.619 \\
6195.433 & 14.0 &  -1.490 & 5.871 \\
6244.466 & 14.0 &  -1.093 & 5.616 \\
5260.387 & 20.0 &  -1.719 & 2.521 \\
5512.980 & 20.0 &  -0.464 & 2.933 \\
5867.562 & 20.0 &  -1.570 & 2.933 \\
6455.598 & 20.0 &  -1.290 & 2.523 \\
$\vdots$ & $\vdots$ & $\vdots$ & $\vdots$ \\
\hline

%% file: nlte.tex
Barium is an important probe of s-process enhancement in stars and  this element is thus very useful to investigate the contribution from different polluter candidates, relative to more r-process dominated species like Eu. However, the derivation of Ba-abundances requires great care, as the lines suffer from hyperfine splitting and isotopic splitting (hereafter HFS) and are very strong at the metallicity of 47Tuc, making them prone to NLTE effects that can be quite significant (see \eg\ \citealt{short,andrievskii}). To take all of the above into account, we performed a full NLTE synthesis for all Ba lines, using the code {\tt MULTI} \citep{carlsson,korotin}, with the line data from \citet{andrievskii}.Three Ba\,{\sc ii} lines are available in our program spectra for the abundance analysis: 5853.68\,\AA, 6141.71\,\AA, and 6496.91\,\AA. Individual line profile fits were made for each of the three lines, allowing for a variation in the macroturbulence in order to ensure the best possible match between the synthesis and the observations. The macroturbulence was treated as a Gaussian broadening of the spectral lines. All barium lines used here, are to some extent blended with lines of iron in particular. The effect is rather significant for the 6141.71\,\AA, and 6496.91\,\AA\ lines. To solve this problem, we fold the NLTE (\texttt{MULTI}) calculations into the LTE synthetic spectrum code {\tt SYNTHV} \citep{tsymbal}, which enable us to calculate synthetic spectra for each Ba\,{\sc ii} line region taking into account all the lines in each region listed in the VALD database\footnote{\url{http://ams.astro.univie.ac.at/vald/}}. For the barium lines, the corresponding departure coefficients (so-called $b$-factors: $b = n_{\rm i}/n^{*}_{\rm i}$ - the ratio of NLTE-to-LTE level populations) are fed into {\tt SYNTHV}, where they are used in the calculation of the line source function and barium line profiles. 

In all cases we were able to produce excellent matches between the calculated and observed spectra (see the example shown in Fig.~\ref{basynth}). In one case (star 6798), we could not use the 6141\,\AA\ line due to strong atmospheric emission, therefore only two lines were used for this star. The size of the NLTE correction varied from star to star and line to line, but reached levels as high as 0.5\,dex in some cases, which is clearly non-negligible. The individual macroturbulence values used for the lines in each star were found to be in excellent agreement with each other, typically differing by less than 0.3\,\kms, with the largest difference found to be 0.5\,\kms. 

That the NLTE corrections are indeed important is further illustrated in Fig.~\ref{bacorr}, where we plot [Ba/Fe] vs. \teff, showing both the LTE and NLTE results. If NLTE is not taken into account, a clear correlation between abundance and \teff\ is observed, which would suggest a problem with the atmospheric models, whereas this behaviour vanishes when NLTE is considered in the abundance derivation.

It is known that aluminium lines are affected by strong NLTE effects (see, for instance, \citealt{gehren}), especially in metal-deficient stars with reduced electron concentration in their atmospheres. To derive NLTE aluminium abundances we used the Al\,I lines at 5557.06\,\AA, 6696.02\,\AA, and 6698.67\,\AA. Our Al atomic model is described in detail in \citet{andrievsky2}. The same method of the blending line treatment, as was used for barium lines, has been applied for the aluminium lines listed above.  It should be noted that the NLTE corrections strongly depend on the stellar parameters. This is illustrated for the reader in Fig.~\ref{al-nlte-vs-teff}. The atomic data for aluminium was taken from \citet{buurman}.

%% file: hfs-lines-sorted-cut.tex
6210.604 & 21.0 &  -2.738 & 0.000 \\
6210.617 & 21.0 &  -2.260 & 0.000 \\
6210.645 & 21.0 &  -2.135 & 0.000 \\
6210.660 & 21.0 &  -2.755 & 0.000 \\
6210.681 & 21.0 &  -2.249 & 0.000 \\
6210.700 & 21.0 &  -2.140 & 0.000 \\
6239.408 & 21.0 &  -2.274 & 0.000 \\
6276.295 & 21.0 &  -2.605 & 0.021 \\
6344.805 & 21.0 &  -3.060 & 0.000 \\
6378.807 & 21.0 &  -2.420 & 0.000 \\
5526.790 & 21.1 &   0.024 & 1.768 \\
5641.001 & 21.1 &  -1.131 & 1.500 \\
5657.896 & 21.1 &  -0.603 & 1.507 \\
5667.149 & 21.1 &  -1.309 & 1.500 \\
5669.042 & 21.1 &  -1.200 & 1.500 \\
5684.202 & 21.1 &  -1.074 & 1.507 \\
$\vdots$ & $\vdots$ & $\vdots$ & $\vdots$ \\
\hline

%% file: abundance_ratio_diff_v7.tex
$\Delta$[O/Fe]    &  $-$0.02 &  0.05 & $-$0.04 &  0.05 & $-$0.02 &  0.04 & $-$0.03 &  0.01 &  0.07 \\
$\Delta$[Na/Fe]   &  $-$0.09 &  0.12 &  0.05 & $-$0.02 & $-$0.01 &  0.04 &  0.03 & $-$0.06 &  0.12 \\
$\Delta$[Mg/Fe]   &   0.03 &  0.01 &  0.02 &  0.02 & $-$0.01 &  0.06 &  0.04 & $-$0.04 &  0.06 \\
$\Delta$[Al/Fe]   &  $-$0.04 &  0.06 &  0.01 & $-$0.01 & 0.00 & $-$0.03 &  0.02 & $-$0.01 &  0.06 \\
$\Delta$[Si/Fe]   &   0.08 & $-$0.07 & $-$0.02 &  0.03 & $-$0.01 &  0.03 & 0.00 & 0.00 &  0.08 \\
$\Delta$[Ca/Fe]   &  $-$0.09 &  0.10 &  0.06 & $-$0.05 &  0.03 & $-$0.02 &  0.02 & $-$0.03 &  0.12 \\
$\Delta$[Sc\,I/Fe]  &  $-$0.10 &  0.18 & 0.00 & $-$0.02 & 0.00 &  0.02 &  0.04 & 0.00 &  0.14 \\
$\Delta$[Sc\,II/Fe] &  $-$0.17 &  0.14 & $-$0.03 & $-$0.04 &  0.01 & $-$0.04 &  0.01 & $-$0.07 &  0.17 \\
$\Delta$[Ti\,I/Fe]  &  $-$0.12 &  0.15 &  0.04 & $-$0.02 &  0.03 & 0.00 &  0.02 & $-$0.02 &  0.14 \\
$\Delta$[Ti\,II/Fe] &  $-$0.11 &  0.13 &  0.04 & $-$0.04 &  0.04 & $-$0.04 &  0.02 & $-$0.03 &  0.14 \\
$\Delta$[V/Fe]    &  $-$0.14 &  0.13 & 0.00 &  0.02 &  0.01 & $-$0.02 &  0.02 & $-$0.02 &  0.14 \\
$\Delta$[Cr/Fe]   &  $-$0.08 &  0.10 &  0.04 & $-$0.03 & 0.00 &  0.02 &  0.02 & $-$0.03 &  0.10 \\
$\Delta$[Fe\,I/H]  &  0.00 & $-$0.02 & $-$0.04 &  0.03 &  0.02 & $-$0.04 & $-$0.03 &  0.03 &  0.06 \\
$\Delta$[Fe\,II/H] &   0.16 & $-$0.18 & $-$0.12 &  0.12 &  0.03 & $-$0.03 & $-$0.06 &  0.07 &  0.22 \\
$\Delta$[Mn/Fe]   &  $-$0.08 &  0.12 &  0.02 &  0.02 &  0.01 &  0.04 &  0.02 & $-$0.04 &  0.11 \\
$\Delta$[Co/Fe]   &  $-$0.02 &  0.03 & $-$0.03 &  0.05 & $-$0.01 &  0.02 & 0.00 &  0.01 &  0.05 \\
$\Delta$[Ni/Fe]   &   0.04 & $-$0.02 & 0.00 &  0.02 &  0.01 &  0.01 &  0.00 &  0.01 &  0.03 \\
$\Delta$[Cu/Fe]   &  $-$0.28 &  0.30 &  0.09 & $-$0.08 &  0.06 & $-$0.04 &  0.08 & $-$0.08 &  0.32 \\
$\Delta$[Zn/Fe]   &   0.10 & $-$0.03 &  0.02 &  0.04 &  0.06 & $-$0.01 &  0.01 & $-$0.02 &  0.08 \\
$\Delta$[Y/Fe]    &  $-$0.15 &  0.19 &  0.03 & $-$0.04 & $-$0.03 &  0.02 &  0.01 & $-$0.03 &  0.18 \\
$\Delta$[Zr/Fe]   &  $-$0.15 &  0.18 &  0.02 &  0.02 &  0.06 & $-$0.01 & $-$0.01 & 0.00 &  0.17 \\
$\Delta$[Mo/Fe]   &  $-$0.12 &  0.12 &  0.01 & $-$0.01 & 0.00 &  0.01 &  0.03 & $-$0.02 &  0.12 \\
$\Delta$[Ru/Fe]   &  $-$0.09 &  0.09 & 0.00 &  0.01 & $-$0.02 &  0.04 &  0.00 & $-$0.01 &  0.10 \\
$\Delta$[Ba/Fe]   &  $-$0.22 &  0.21 &  0.10 & $-$0.06 &  0.04 & $-$0.09 &  0.03 & 0.00 &  0.24 \\
$\Delta$[La/Fe]   &  $-$0.18 &  0.19 &  0.03 & $-$0.04 & $-$0.03 &  0.02 & 0.00 & $-$0.01 &  0.19 \\
$\Delta$[Ce/Fe]   &  0.00 &  0.34 &  0.20 &  0.12 &  0.15 &  0.17 &  0.01 & $-$0.02 &  0.28 \\
$\Delta$[Pr/Fe]   &  $-$0.19 &  0.22 &  0.05 & $-$0.05 & $-$0.01 &  0.01 &  0.01 & $-$0.02 &  0.21 \\
$\Delta$[Nd/Fe]   &  $-$0.20 &  0.23 &  0.05 & $-$0.06 & $-$0.01 &  0.01 &  0.03 & $-$0.04 &  0.22 \\
$\Delta$[Eu/Fe]   &  $-$0.13 &  0.16 &  0.04 & $-$0.03 & $-$0.02 &  0.03 &  0.01 & $-$0.03 &  0.15 \\
$\Delta$[Dy/Fe]   &  $-$0.23 &  0.24 &  0.05 & $-$0.07 & $-$0.03 &  0.01 &  0.04 & $-$0.06 &  0.25 \\

%% file: fund_params.tex
ID  & \teff\ & \logg & $\xi_t$ & \feh\ & V$_{{\rm macro}}$\\
\hline\hline
  \textbf{1062} & 3870 & 0.45 & 1.30 & -0.78 &6.58 \\
  4794 & 4070 & 1.15 & 1.30 & -0.66 &5.82 \\
  \textbf{5265} & 3870 & 0.30 & 1.25 & -0.69 &6.14 \\
  5968 & 3970 & 0.85 & 1.40 & -0.79 &5.92 \\
  6798 & 4000 & 0.90 & 1.30 & -0.69 &5.63 \\
 10237 & 4280 & 1.20 & 1.60 & -0.83 &5.93 \\
 13396 & 4190 & 1.45 & 1.60 & -0.83 &5.35 \\
 20885 & 4260 & 1.35 & 1.90 & -0.84 &5.60 \\
 \textbf{27678} & 3870 & 0.35 & 1.20 & -0.76 &6.59 \\
 \textbf{28956} & 3900 & 0.30 & 1.60 & -0.86 &6.94 \\
 29861 & 4160 & 1.20 & 1.50 & -0.84 &5.28 \\
 \textbf{38916} & 4080 & 0.85 & 1.40 & -0.83 &5.67 \\
 \textbf{40394} & 3890 & 0.45 & 1.10 & -0.71 &6.88 \\
\hline

%% file: abundance-ratios-cut-v7.tex
ID & [Fe\,I/H] & $\sigma_{{\rm tot}}$ & N & [Fe\,II/H] & $\sigma_{{\rm tot}}$ & N & [O/Fe] & $\sigma_{{\rm tot}}$ & N &  [Na/Fe] & $\sigma_{{\rm tot}}$ & N & [Mg/Fe] & $\sigma_{{\rm tot}}$ & N \\
\hline\hline
  \textbf{1062} & $-$0.78 & 0.06 & 44 & $-$0.80 & 0.22 & 10 &  0.23 & 0.10 &  2 &  0.24 & 0.14 &  2 &  0.46 & 0.07 &  2 \\
  4794 & $-$0.66 & 0.06 & 44 & $-$0.66 & 0.22 & 10 &  0.38 & 0.10 &  2 &  0.15 & 0.14 &  2 &  0.40 & 0.09 &  2 \\
  \textbf{5265} & $-$0.69 & 0.06 & 43 & $-$0.69 & 0.22 & 11 &  0.00 & $-$    &  2 &  0.31 & 0.16 &  2 &  0.40 & 0.07 &  3 \\
  5968 & $-$0.79 & 0.06 & 49 & $-$0.79 & 0.23 &  7 &  0.41 & 0.10 &  2 &  0.08 & 0.14 &  2 &  0.49 & 0.08 &  2 \\
  6798 & $-$0.69 & 0.06 & 44 & $-$0.67 & 0.22 & 14 &  0.37 & $-$    &  2 &  0.01 & 0.13 &  2 &  0.44 & 0.09 &  2 \\
 10237 & $-$0.83 & 0.06 & 42 & $-$0.83 & 0.22 & 13 &  0.57 & 0.10 &  2 &  0.04 & 0.13 &  2 &  0.32 & 0.09 &  2 \\
 13396 & $-$0.83 & 0.06 & 47 & $-$0.86 & 0.22 & 11 &  0.48 & 0.10 &  1 &  0.07 & 0.13 &  2 &  0.44 & 0.07 &  3 \\
 20885 & $-$0.84 & 0.06 & 37 & $-$0.86 & 0.22 & 14 &   $-$   & $-$    &  0 &  0.11 & 0.13 &  2 &  0.47 & 0.09 &  3 \\
\textbf{27678} & $-$0.76 & 0.06 & 47 & $-$0.76 & 0.22 & 12 &  0.13 & 0.10 &  2 &  0.50 & 0.14 &  2 &  0.45 & 0.10 &  2 \\
\textbf{28956} & $-$0.86 & 0.06 & 46 & $-$0.82 & 0.22 & 13 &  0.06 & 0.12 &  2 &  0.41 & 0.13 &  2 &  0.46 & 0.07 &  3 \\
 29861 & $-$0.84 & 0.06 & 40 & $-$0.82 & 0.22 & 12 &  0.47 & 0.10 &  2 &  0.10 & 0.13 &  2 &  0.40 & 0.11 &  1 \\
\textbf{38916} & $-$0.83 & 0.06 & 48 & $-$0.85 & 0.22 & 14 &   $-$   & $-$    &  0 &  0.42 & 0.14 &  2 &  0.52 & 0.07 &  3 \\
\textbf{40394} & $-$0.71 & 0.06 & 43 & $-$0.67 & 0.22 & 10 &  0.23 & 0.10 &  2 &  0.26 & 0.14 &  2 &  0.43 & 0.08 &  3 \\
\hline

%% file: abundancedispersion_IQR_v2.tex
Elem. & IQR & Median  & $\bar{\sigma}$ \\
\hline\hline
 {[Fe~I/H]} &  0.12 & -0.79 & 0.06 \\
{[Fe~II/H]} &  0.14 & -0.80 & 0.22 \\
   {[O/Fe]} &  0.26 &  0.30 & 0.10 \\
  {[Na/Fe]} &  0.23 &  0.21 & 0.14 \\
  {[Mg/Fe]} &  0.06 &  0.44 & 0.08 \\
  {[Al/Fe]} &  0.11 &  0.21 & 0.08 \\
  {[Si/Fe]} &  0.05 &  0.32 & 0.09 \\
  {[Ca/Fe]} &  0.12 &  0.24 & 0.13 \\
{[Sc~I/Fe]} &  0.19 &  0.01 & 0.15 \\
{[Sc~II/Fe]} & 0.06 &  0.21 & 0.17 \\
{[Ti~I/Fe]} &  0.08 &  0.33 & 0.15 \\
{[Ti~II/Fe]} & 0.16 &  0.41 & 0.15 \\
   {[V/Fe]} &  0.12 &  0.17 & 0.14 \\
  {[Cr/Fe]} &  0.08 & -0.03 & 0.11 \\
  {[Mn/Fe]} &  0.08 & -0.20 & 0.13 \\
  {[Co/Fe]} &  0.10 & -0.00 & 0.07 \\
  {[Ni/Fe]} &  0.03 & -0.12 & 0.04 \\
  {[Cu/Fe]} &  0.18 & -0.14 & 0.35 \\
  {[Zn/Fe]} &  0.05 &  0.26 & 0.13 \\
   {[Y/Fe]} &  0.20 &  0.07 & 0.19 \\
  {[Zr/Fe]} &  0.16 &  0.41 & 0.17 \\
  {[Mo/Fe]} &  0.07 &  0.55 & 0.13 \\
  {[Ru/Fe]} &  0.09 &  0.51 & 0.13 \\
  {[Ba/Fe]} &  0.09 &  0.25 & 0.24 \\
  {[La/Fe]} &  0.21 &  0.07 & 0.21 \\
  {[Ce/Fe]} &  0.13 & -0.04 & 0.32 \\
  {[Pr/Fe]} &  0.09 & -0.04 & 0.24 \\
  {[Nd/Fe]} &  0.09 &  0.04 & 0.24 \\
  {[Eu/Fe]} &  0.22 &  0.32 & 0.19 \\
  {[Dy/Fe]} &  0.15 &  0.70 & 0.27 \\
\hline